\documentclass[]{pasj01}

\usepackage{multirow}
\usepackage{natbib}

\def\aaps{{A\&A Suppl.}}
\def\apj{{ApJ}}
\def\mnras{{MNRAS}}
\def\aap{{A\&A}}
\def\pasp{{PASP}}
\def\pasj{{PASJ}}
\def\icarus{{Icarus}}
\def\nat{{Nature}}
\def\aj{{AJ}}
\def\apjs{{ApJS}}

\begin{document} 
\Received{}
\Accepted{}

\title{AKARI/IRC Near-Infrared Asteroid Spectroscopic Survey: AcuA-spec}

\author{Fumihiko \textsc{Usui}\altaffilmark{1}}
\altaffiltext{1}{%
Center for Planetary Science, Graduate School of Science, Kobe University, %
7-1-48, Minatojima-Minamimachi, Chuo-Ku, Kobe 650-0047, Japan}
\email{usui@cps-jp.org}

\author{Sunao \textsc{Hasegawa}\altaffilmark{2}}
\altaffiltext{2}{%
Institute of Space and Astronautical Science, Japan Aerospace Exploration Agency, %
3-1-1 Yoshinodai, Chuo-ku, Sagamihara 252-5210, Japan}

\author{Takafumi \textsc{Ootsubo}\altaffilmark{2}}

\author{Takashi \textsc{Onaka}\altaffilmark{3}}
\altaffiltext{3}{%
Department of Astronomy, Graduate School of Science, The University of Tokyo, %
7-3-1, Hongo, Bunkyo-ku, Tokyo 113-0033, Japan}


\maketitle

\begin{abstract}
Knowledge of water in the solar system is important for understanding of a wide range of 
evolutionary processes and the thermal history of the solar system. To explore 
the existence of water in the solar system, it is indispensable to investigate 
hydrated minerals and/or water ice on asteroids. These water-related materials 
show absorption features in the 3-$\micron$ band (wavelengths from 2.7 to 3.1~$\micron$). 
We conducted a spectroscopic survey of asteroids in the 3-$\micron$ band 
using the Infrared Camera (IRC) on board the Japanese infrared satellite AKARI.
In the warm mission 
period of AKARI, 147 pointed observations were performed for 66 asteroids in the grism mode 
for wavelengths from 2.5 to 5~$\micron$. According to these observations, most C-complex asteroids 
have clear absorption features ($> 10\%$ with respect to the continuum) 
related to hydrated minerals at a peak wavelength of approximately 
2.75~$\micron$, while S-complex asteroids have no significant feature in this wavelength range.
The present data are released to the public as the Asteroid Catalog using AKARI Spectroscopic Observations (AcuA-spec). 
\end{abstract}

\section{Introduction}
Water is found in various forms in our solar system and is one of the most important ingredients in the origin of life. It also has vital implications on the exploration of 
extrasolar planets and provides evidence for the evolution of the solar system, especially 
its thermal history. 
Silicate minerals account for a large fraction of solid materials in the solar system 
and water exists as ice in small solar system bodies beyond Jupiter. 
Hydrated minerals (any mineral that contains H$_2$O or OH) are formed 
in environments where anhydrous rock and liquid water exist together 
with a certain pressure and temperature, resulting from aqueous alteration
(e.g., \citealt{Brearley2006} and references cited therein). 
Because hydrated minerals are stable even above the sublimation temperature of water ice, 
they become an important tracer of water
present in the history of the solar system unless they were reset by a temperature change after formation. 
The study of hydrated minerals is therefore important for understanding of the origin of Earth's water 
and unravelling of the earliest thermal processes in the solar system.
Most asteroids have not experienced sufficient thermal evolution to differentiate into 
layered structures like terrestrial planets since their formation; thus, 
asteroids are considered to record the initial conditions of our solar nebula of 4.6 Ga ago. 
To explore the existence of water in the present solar system, it is necessary to investigate 
the presence of hydrated minerals and water ice on various types of asteroids. 

Meteorites collected 
on the Earth have typically fallen from asteroids
(e.g., \citealt{Morbidelli1998}) and bring useful 
information for asteroid research. However, it is difficult to correctly measure the water 
content because meteorites have been contaminated by atmospheric water 
(e.g., \citealt{Beck2010,2015Icar..257..185T,Mogi2017}). 
Observations with astronomical telescopes are needed to investigate asteroids without 
terrestrial alteration. 
Hydrated minerals and water ice exhibit diagnostic absorption features in the so-called 3-$\micron$ band
(approximately 2.5--3.5~$\micron$ wavelength range, \citealt{Rivkin2002}). 
Features at $\sim$ 2.7~$\micron$ are attributed to hydrated minerals and those at $\sim$ 3.05~$\micron$ 
to water ice. 
Other materials also display spectral features at these wavelengths, such as ammonium-bearing 
phyllosilicate~\citep{King1992}, 
mineral brucite (magnesium hydroxide; \citealt{Beck2011}), 
and absorbed water molecules in regolith particles (e.g., in the lunar rocks or soils; \citealt{Clark2009a}). 
Many spectroscopic surveys have been conducted in the 3-$\micron$ band using ground-based observatories. 
\cite{2012Icar..219..641T} observed 28 outer main-belt asteroids
with the NASA Infrared Telescope Facility (IRTF) on the summit of Mauna Kea, Hawaii, 
and classified them into four spectral groups
based on the absorption shapes: 
sharp, rounded, Ceres-like, and Europa-like, and discussed the distribution and abundance of hydrated 
minerals in the outer main-belt region. 
However, space-borne telescopes can perform more accurate 
observations for the identification of mineral species, 
because the spectrum observed with a ground-based telescope (2.5-2.85~$\micron$) is severely affected 
by telluric absorption (e.g., \citealt{Rivkin2002}).

AKARI~\citep{Murakami2007}, launched in 2006, is a Japanese satellite mission 
fully dedicated to a wide range of infrared astronomy, including 
galaxy evolution, stellar formation and evolution, interstellar media, and solar system objects.
As part of our research, we conducted two types of asteroid survey with AKARI. 
One is a mid-infrared asteroid survey. 
Using the mid-infrared part of all-sky survey data 
obtained with the Infrared Camera (IRC; \citealt{Onaka2007,Ishihara2010}) on board AKARI, 
we constructed a size and albedo catalog of 5120 asteroids~\citep{Usui2011}, 
which is summarized in the Asteroid Catalog using AKARI (AcuA)\footnote{%
This catalog, as well as the infrared flux data of individual asteroids~\citep{AliLagoa2018}, is 
open to the public at the following URLs: 
http://darts.isas.jaxa.jp/astro/akari/catalogue/AcuA.html and 
http://www.ir.isas.jaxa.jp/AKARI/Archive/Catalogues/Asteroid\_Flux\_V1/
}. 
This is an unbiased asteroid catalog at two mid-infrared bands (9~$\micron$ and 18~$\micron$), 
which fully covers objects with a diameter of $> 40$~km in the main-belt region~\citep{Usui2014}. 
It was conducted using observations made during the cryogenic phase (Phase 1 and 2). 
The other survey is a near-infrared spectroscopic survey, which is described in this paper. 
Low-resolution spectroscopic observations were performed using the near-infrared channel (2.5--5~\micron) 
of the IRC, which provide valuable data thanks to 
its high sensitivity and unique wavelength coverage~\citep{Ohyama2007}. 
Note that the Infrared Space Observatory (ISO; \citealt{Kessler1996}) has a spectroscopic sensitivity to detect 
only the largest objects in the main-belt region~\citep{1999ESASP.427..165D,1997PhDT........10R} 
and that the Infrared Spectrograph (IRS) of the Spitzer space telescope~\citep{Werner2004} only covers wavelengths longer than 5~$\micron$. 

In this study, we report a spectroscopic survey of asteroids using the IRC on board AKARI. 
During the warm mission period of AKARI (Phase 3; \citealt{Onaka2012}), 
147 pointed observations were performed for 66 asteroids in the grism mode at wavelengths from 2.5 to 5~$\micron$. 
The observed objects comprise C-complex ($\times 23$), S-complex ($\times 17$), X-complex ($\times 22$), 
D-complex ($\times 3$), and V-type ($\times 1$) asteroids, 
all of which are in the main-belt region and have diameters of 40 km or larger. 

This paper is organized as follows: 
section 2 describes the observations of asteroids made with AKARI and the data reduction process used to derive reflectance spectra; 
section 3 presents the characteristics of the obtained spectra; and the results are discussed in section 4. 

\section{Observations and Data Reduction}
\subsection{NIR grism spectroscopic observations with the AKARI/IRC}
\label{sec:NIR grism Spectroscopic observations with the AKARI/IRC}
AKARI was launched on 2006 February 21 UT, and its liquid helium cryogen boiled off 
on 2007 August 26 UT, 550 days after the launch~\citep{Murakami2007}.
This cryogenic phase is referred to as Phase 1 and 2. 
In the post-helium phase (Phase 3), the telescope and its scientific instruments were maintained at approximately 40 K by the mechanical cooler and only near-infrared observations were carried out until 2010 February. 
The main purpose of AKARI was to perform all-sky surveys during Phase 1 and 2 by continuous scanning of the sky at a constant scan speed; i.e., in 
the survey observation mode. 
AKARI also has a capability of a pointed observation, which is occasionally inserted 
into the continuous survey observation (the observational strategy of AKARI is discussed in \citealt{Matsuhara2005}). 
In Phase 3, only pointed observations with IRC were available, 
in which the same attitude control mode was employed as that during Phase 1 and 2 operations. 
This pointed observation mode is employed for deep imaging or spectroscopy. 

\begin{figure}
\begin{center}
\includegraphics[width=140mm]{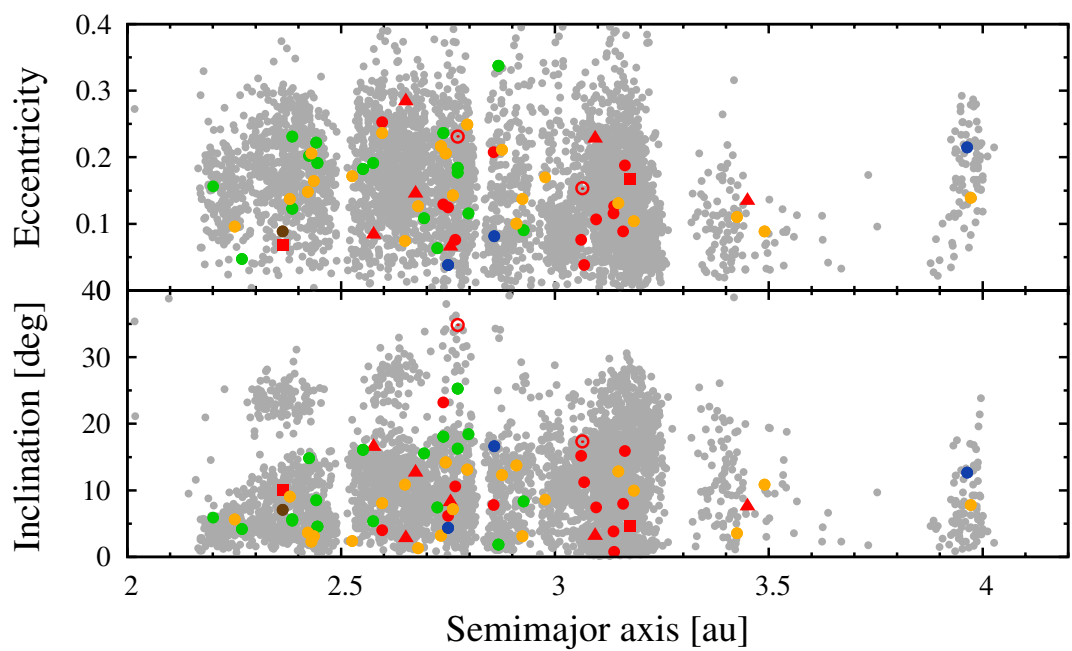}
\end{center}
\caption{
Distribution of the orbital elements (semimajor axis, inclination, and eccentricity) of our target asteroids. 
Red, green, yellow, blue, and brown denote C-, S-, X-, D-complex, and V-type asteroids, respectively. 
C-complex asteroids are subdivided into 
rectangles, triangles, open circles, and filled circles, as 
Cgh-, Ch-, B or Cb, and other C-type, respectively. 
Gray dots are all asteroids of AcuA data~\citep{Usui2011}.}\label{fig:orbital_elements}
\end{figure}

\begin{figure}
\begin{center}
\includegraphics[width=140mm]{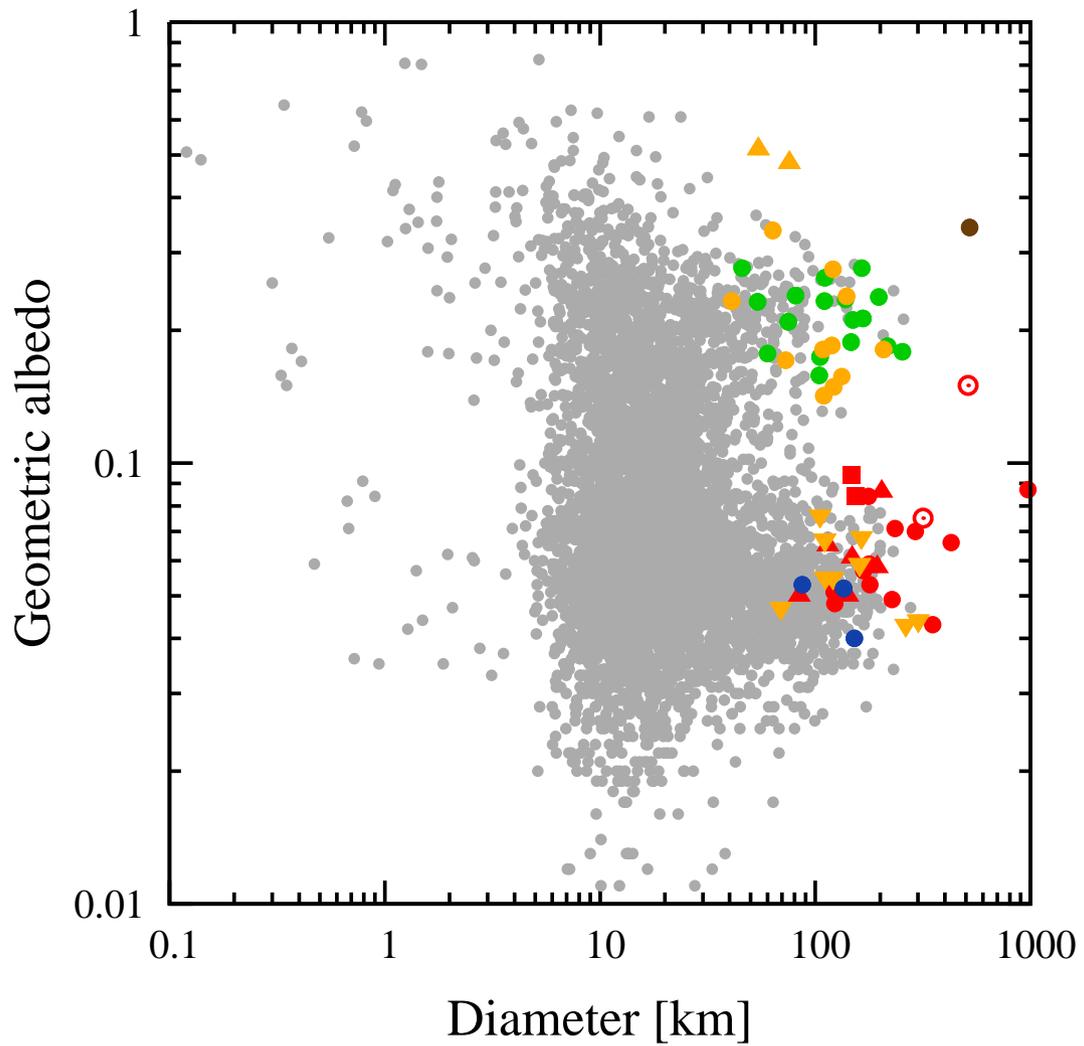}
\end{center}
\caption{Size and geometric albedo distribution of our targets. Colors and marks are the same as in figure~\ref{fig:orbital_elements}.}\label{fig:size-albedo} 
\end{figure}

Our targets were 66 objects selected from main-belt asteroids, the Cybeles, the Hildas, 
and one near-Earth object, 4015~Wilson-Harringtion, according to the visibility from AKARI.
Most of these observations were performed as part of the AKARI Mission Program entitled 
``Origin and Evolution of Solar System Objects'' (SOSOS), while four observations were executed within the framework of 
the AKARI director's time program (DT): 
two observations were made for 6~Hebe and the other two for 128~Nemesis.
The physical properties and taxonomic classification of our targets are summarized in tables~\ref{tab:object_info} 
and \ref{tab:object_tax}. 
All objects have entries of the size and albedo \citep{Usui2011,Mainzer2011}, rotational period \citep{Warner2009}, 
and taxonomic information from visible to near-infrared wavelengths (from 0.4--2.5~$\micron$; hereafter vis-NIR) 
compiled by~\cite{Hasegawa2017}. 
The distribution of the orbital elements and the size and albedo diagrams of our targets are shown in 
figures~\ref{fig:orbital_elements} and \ref{fig:size-albedo}, respectively.

\begin{figure}
\begin{center}
\includegraphics[width=140mm]{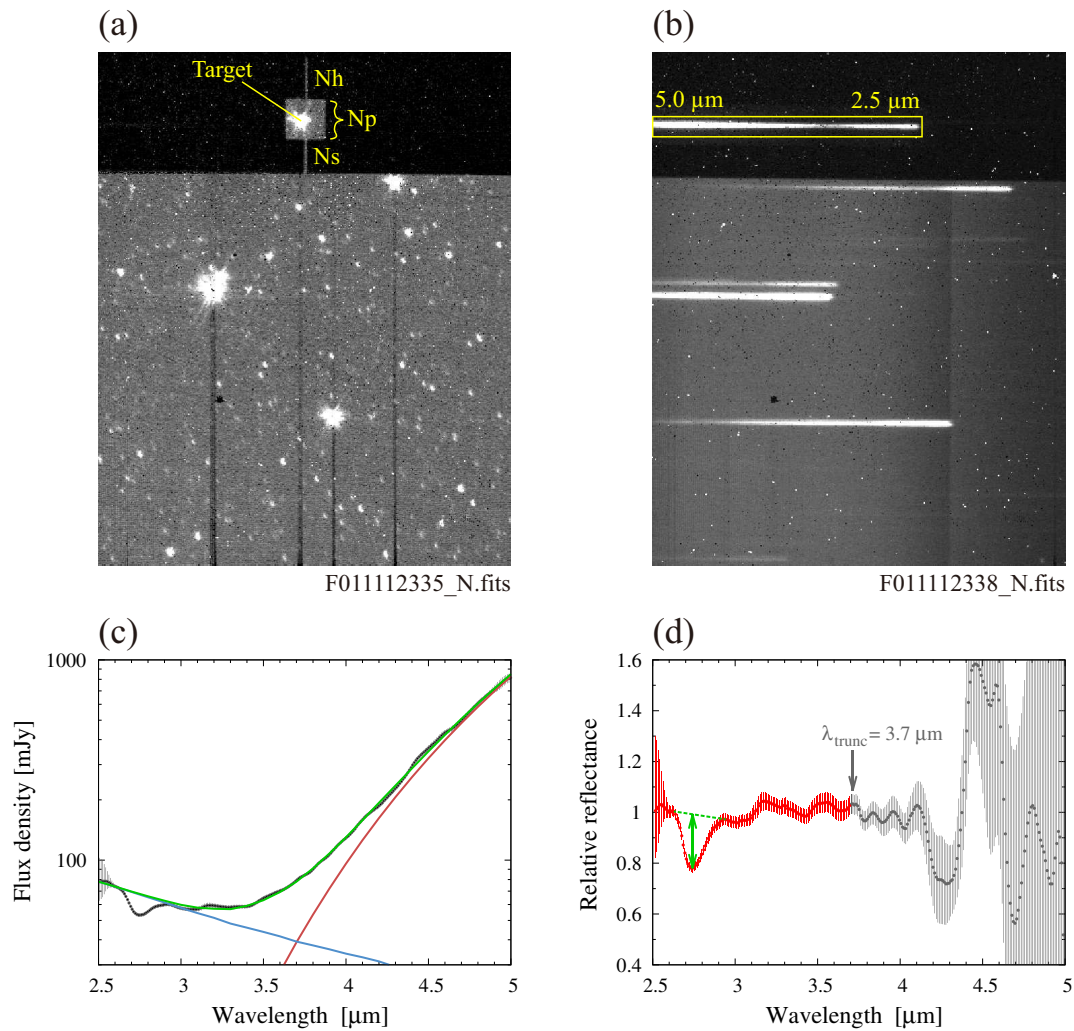}
\end{center}
\caption{
Example of data reduction for C-type asteroid 511~Davida (ObsID:1520065-001, observed on 2008 November 16). 
(a) Reference image in the N3 band, which is used to derive the wavelength reference position in the spectral images.  
(b) Spectroscopic image with the grism (NG). 
One pointed observation consists of four spectroscopic frames, one reference frame, and four or five 
spectroscopic frames (total number of frames depends on the attitude stability of the satellite). 
The target is placed on the $1^{\prime} \times 1^{\prime}$ ``Np'' window to avoid contamination from background stars. 
(c) Spectrum extracted from the target using the toolkit. The red, blue, and green curves denote the thermal component of the spectrum 
calculated by the NEATM that is removed to derive the reflectance spectrum, the spectrum of the reflected sunlight component, and the total modeled spectrum, respectively.
(d) Reflectance spectrum normalized at 2.6~$\micron$. Spectrum beyond the truncated wavelength ($\lambda_\mathrm{trunc}$)
cannot be used due to uncertainty of the thermal model
(see section~\ref{sec:Criterion to reject spectra severely contaminated by thermal emission}). 
The green dotted line denotes the continuum and the arrow shows the point where the band depth is measured. }\label{fig:data-example} 
\end{figure}

Observations were conducted 147 times from 2008 May to 2010 February. Observation logs are given in table~\ref{tab:obslog}. 
Typically, two or three pointed observations were performed for each target, 
whereas three targets (8~Flora, 69~Hesperia, and 4015~Wilson-Harrington) 
were observed only once. 
Most objects were observed within $\sim$ 100~min intervals, which were the orbital period of the satellite. 
Exceptions are 216~Kleopatra and 704~Interamnia, both of which were observed at intervals of more than 14 months because of the scheduling constraint, but not for any scientific reason. 
The direction of AKARI observations was severely limited to within 
90$^\circ$ $\pm$ 0.7$^\circ$ of the solar elongation angle 
due to the design of the orbit and attitude control system to avoid 
radiation from the Earth and the Sun. 
Within this geometric constraint, 
our targets in the main-belt region were located between phase angles of 15$^\circ$ and 32$^\circ$. 
We made observations with the AKARI Astronomical Observation Template (AOT) of spectroscopy dedicated to Phase 3 
(IRCZ4; see IRC Data User Manual~\footnote{http://www.ir.isas.jaxa.jp/AKARI/Observation/support/IRC/}), 
which is equivalent to IRC04 in Phase 1 and 2 \citep{Onaka2007, Ohyama2007}. 
The grism mode covered a wavelength range of 2.5--5~$\micron$ with 
a spectral resolution of $R \sim 100$. 
Within a single AOT operation lasting approximately 10 min, 8 or 9 spectroscopic images were taken with the grism, 
as well as a photometric image (called a reference image) through 
a broadband filter in the wavelength range of 2.7--3.8~$\micron$ centered at 3.2~$\micron$ (the N3 band). 
The effective exposure time for each frame was 44.41~s 
(see IRC Data User Manual). 
Total exposure time of one pointed observation is 310--400~sec depending on the attitude stability of the satellite; 
thus our observations provided a flux-limited sample. 
The targets were put on the $1^{\prime} \times 1^{\prime}$ aperture mask 
(the ``Np'' window; \citealt{Onaka2007}, also see figure~\ref{fig:data-example}) to minimize contamination 
from nearby objects (e.g., \citealt{Ootsubo2012}). 
AKARI was also equipped with narrow slits (``Nh'' of width of $3^{\prime\prime}$ and 
``Ns'' of width of $5^{\prime\prime}$); however, these modes were designed for diffuse sources. 
The absolute accuracy of the telescope pointing of the satellite is a few arcseconds; thus, we used the ``Np'' window instead of ``Nh'' or ``Ns'' slits for point sources. 
For the ``Np'' observations, it was not necessary to consider slit loss because of the large width of this mask. 
Note that the full width at the half maximum of the point spread function (PSF) 
in the N3 band is 3.2 pixels or 
4\farcs{67} 
(\citealt{Onaka2008,Onaka2010}; also see IRC Data User Manual). 

\subsection{Data reduction}
\label{sec:Data Reduction}
Figure~\ref{fig:data-example} shows an example of data reduction for C-type asteroid 511~Davida (ObsID:1520065-001).  

\subsubsection{Extraction of one-dimensional spectra}
\label{sec:Extraction of one-dimensional spectra}
The raw data were reduced using the IDL-based software package in the IRC Spectroscopy Toolkit for Phase 3 data, version 20170225RC (\citealt{Ohyama2007,Usui2018}; also see IRC Data User Manual). Standard array image processing, such as dark subtraction, cosmic-ray removal, linearity correction, 
and various image anomaly corrections 
(bad-pixel masking, correction of tilt of the spectrum, and the first-order sky subtraction), 
were first performed for each exposure frame 
with the toolkit. 
Then, multiple-exposure frames were added, taking into 
account the relative image shift due to the satellite attitude drift among the frames.
Because AKARI does not have a tracking mode for moving objects, the resulting two-dimensional spectra 
with the standard toolkit was blurred due to motion of the object. 
Note that the velocities of apparent motion of the targets 
were typically less than 
7\farcs{5} / 10~min (or $\sim$~5 pixel / 10~min); thus, 
the targets were located within the field of view of the ``Np'' window during the observations 
(with the exception of 4015~Wilson-Harrington, which was 30$^{\prime\prime}$ / 10 min). 
The movements of the target asteroid during the observation of each frame were calculated to add 
to the relative image shift, and multiple-exposure frames were shift-and-added 
to extract two-dimensional spectral image of the objects. 
This function was implemented in the toolkit for the data of solar system objects 
(details are given in \citealt{Ootsubo2012}). 
Using the combination of shift-and-add with the movement of the asteroid and the 3-$\sigma$ clipping methods, 
the effects of hot pixels were reduced in the stacked images, although the number of hot pixels in Phase 3 increased compared with 
Phase 1 and 2. 
The fluctuation of the background signals causes a random noise of the flux uncertainties of the target spectra. 
With the standard toolkit, 
the sky background was estimated from the adjacent region of the target in the ``Np'' window 
and subtracted from the stacked images. 
In some cases, however, the background signals are not able to be estimated properly 
from the adjacent region because 
of contamination of neighboring stars in the ``Np'' window.  
In this study, 
wavelength-dependent sky background 
was 
estimated from 
the diffuse sky spectrum of the ``Ns'' slit for the same pointed observation 
which was checked as contamination-free ``blank'' sky by visual inspection. 

The absolute flux calibration (spectral response calibration) of the AKARI/IRC spectroscopy is based on 
the observation of active galactic nuclei (AGNs) and spectroscopic standard stars 
obtained with the same instrumental setup. The AGNs without significant 
emission or absorption features are used as calibrators of redder objects
(contributed in longer wavelengths), 
and the standard stars whose spectra are already templated are used as bluer objects
(contributed in shorter wavelengths). 
13 featureless AGNs and six K- and A-type standard stars were selected as 
flux templates. The observations of these objects were performed 
within the framework of the DT program, separately from individual observations. 
Comparing the modeled templates of the objects 
with the observed flux count in analog-to-digital units (ADU), the spectral response
was obtained as a function of wavelength. 
This spectral response was implemented in the toolkit. 
Details are given in \cite{Baba2016} and Baba et al., submitted. 

In general, the spectroscopic flat-fields are made by gathering a large number of blank sky spectroscopy images that are
combined and normalized so that any faint object spectra are removed by clipping averaging 
techniques. For the AKARI observations, flat data were not taken in individual pointed observations because 
the observational time was severely limited due to the avoidance constraints of the satellite. 
As the natural background is faint in the near-infrared for small-aperture spectroscopy of the IRC (Np, Ns, and Nh), 
the quality of the flat data cannot be improved 
substantially even after stacking of multi-pointed observational data. 
Thus, the quality of the processed spectra is, unfortunately, limited by 
the quality of the flat data, not by the dark current nor photon noise. 
If the spectra have low S/N ratios, then applying a flat could degrade the data. 
Therefore, we skipped performing flat fielding in the data reduction process 
(with an option called {\tt /no\_slit\_flat} prepared to disable 
the flat correction in the toolkit: see IRC Data User Manual). 

It is reported that the sensitivity decreased by a maximum of approximately 10\% during Phase 3 
with the increase of the IRC detector temperature~\citep{Onaka2010}. 
This can be approximated by a linear function of the detector temperature $T$ 
(Baba et al., submitted) as: 
\begin{eqnarray}
f(T) &=& 1 + a (T-T_0)\ ,
\end{eqnarray}
where $f(T)$ is the correction factor, $a = -0.0290$~K$^{-1}$, and $T_0 = 43.55$~K. 
For ObsID:1521116-001 observed on 
2009 November 18, the recorded detector temperature was 
the highest ($T = 45.13$~K) among all asteroid observations during Phase 3, 
and the correction factor was given as $f(T) = 0.954$. 
Then the flux density is scaled by $1/f(T) = 1.048$ to obtain the corrected value. 
This correction factor is assumed to not have wavelength-dependency. 
In the spectroscopic analysis, the flux and the wavelength accuracy strongly depend on the accuracy of the wavelength zero point, which was estimated on the reference image of the N3 band. It was estimated to be, at worst, 1 pixel in our analysis. Thus, the flux and the wavelength uncertainties were estimated by calculating how much the spectrum changed when the wavelength zero point was shifted by $\pm 1$ pixel~\citep{Shimonishi2013}.

Finally, one-dimensional spectra were extracted from the two-dimensional images by summing signals 
over 7~pixels (approximately 10\farcs{5}) 
in the spatial direction to reduce the effect of hot pixels and/or cosmic rays that hit the detector. 
%
Then, one-dimensional spectra were extracted. 
The obtained one-dimensional spectra were smoothed along their wavelength with 5 pixels (i.e., $1.5 \times$ PSF) for further analyses. 
In this study, we focus on broader features  (bandwidth of $> 0.1~\micron$) appeared 
at around 2.7~$\micron$ and 3.1~$\micron$ in the spectra (c.f., table~1 in \citealt{Rivkin2015AIV}). 
There are wavy patterns in other wavelengths remaining in some extracted spectra,
which have not seen in meteorite spectra. It is likely that most of them are spurious due 
to the contamination of neighboring stars or insufficient background subtraction, 
although we cannot completely rule out that they are real features. 
In the following, we 
do not discuss other features. 
The obtained spectra of each pointed observation are summarized in supplementary data. 

After visual inspection, it was found that 
some observational data were contaminated by ghost patterns~\citep{Egusa2016} or by
nearby stars that happened to be in the same field of the ``Np'' window. 
The data of the following observations were therefore removed from further analyses (see table~\ref{tab:obslog}); 
ObsID:1521233-001 for 9~Metis, 
ObsID:1520189-001 for 44~Nysa, 
ObsID:1521191-001 for 49~Pales, 
ObsID:1521212-001 for 145~Adeona, and 
ObsID:1520161-001 for 250~Bettina. 

\subsubsection{Instrument linearity and saturation}
\label{sec:saturation}
Saturation of the observed signal makes 
severe problems for 
larger asteroids, which need to be handled with care. 
The saturation level of the detector during Phase 3 observations was reported as 2000 ADU per pixel 
(\citealt{Onaka2008}; see also IRC Data User Manual), which roughly corresponds to $\sim$ 1~Jy for the grism mode. 
A brighter observed flux than this value cannot be assured in the linearity correction of the detector. 
This was the case for the largest objects: 1~Ceres, 2~Pallas, and 4~Vesta. 
In this work, the following empirical method was used to determine the available wavelength range and extract the spectra. 

\begin{figure}
\begin{center}
(a) 1~Ceres\\
\includegraphics[width=140mm]{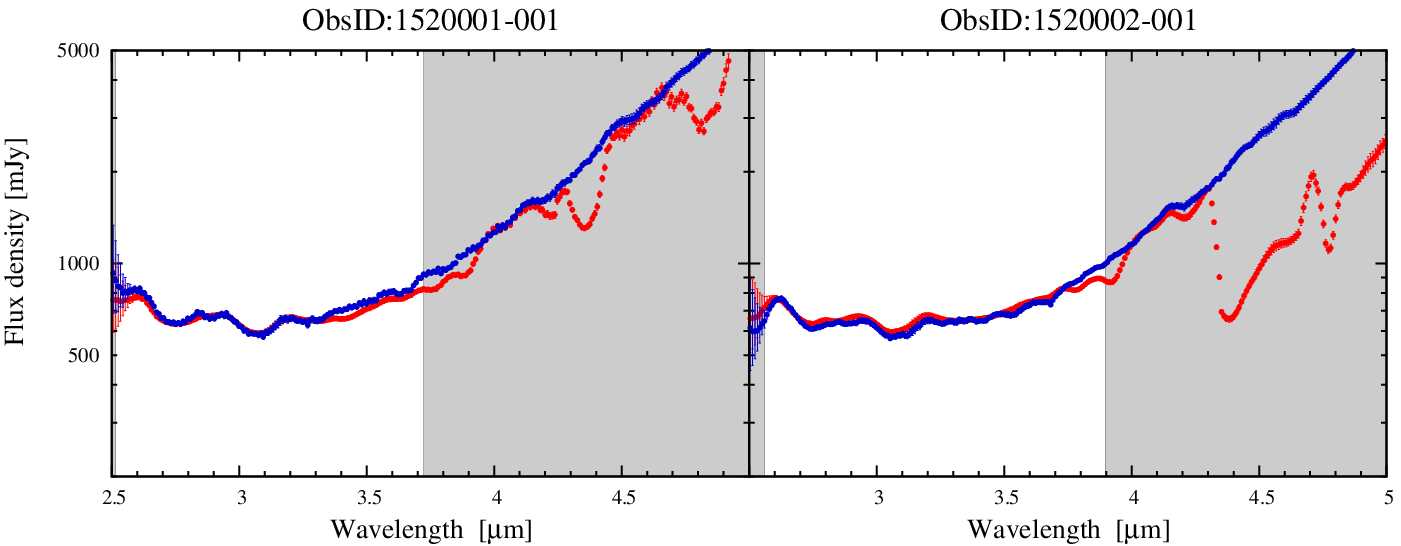}\\
(b) 2~Pallas\\
\includegraphics[width=140mm]{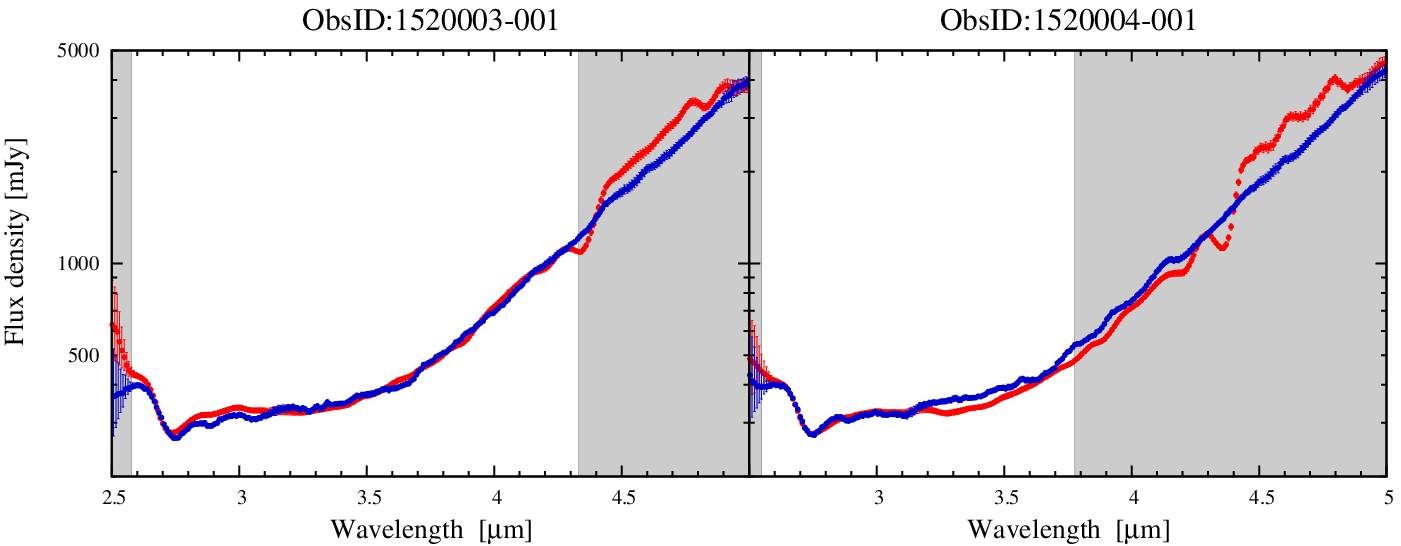}\\
(c) 4~Vesta\\
\includegraphics[width=140mm]{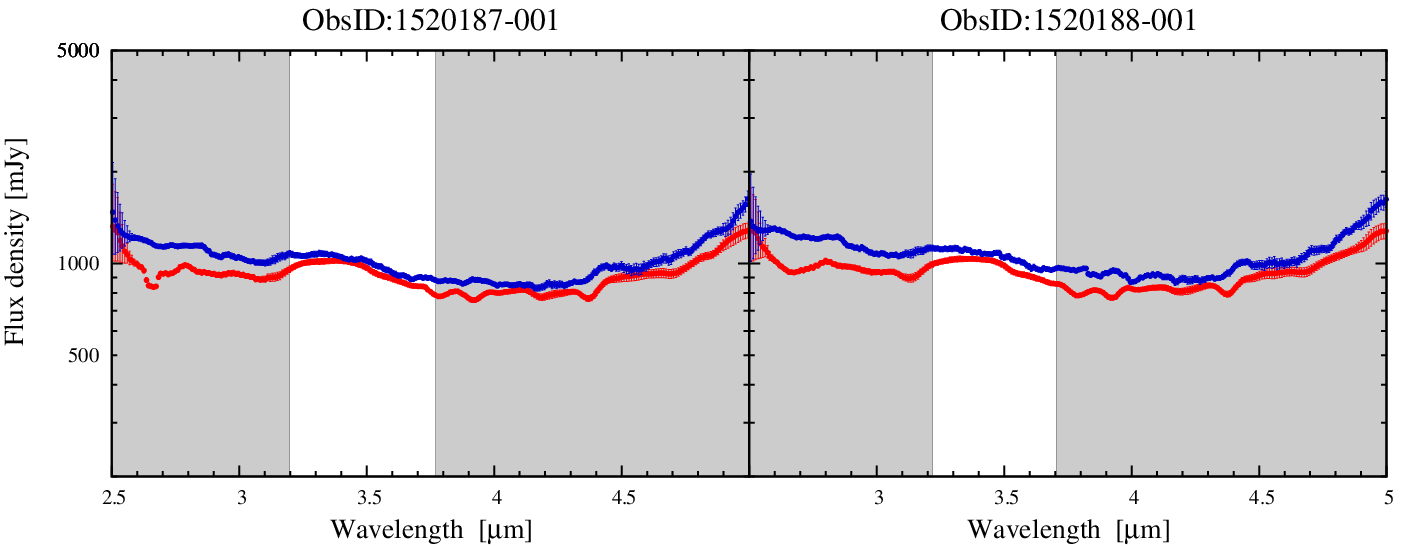}
\end{center}
\caption{
AKARI IRC 2.5-5.0~$\micron$ spectra of 1~Ceres, 2~Pallas, and 4~Vesta of each pointed observation. 
The red and blue dots denote the long exposure and short exposure data, respectively. 
The gray region denotes the unreliable wavelength range determined by inconsistency between the long and short exposure data.
}\label{fig:spectra-largest}
\end{figure}

The minimum observation unit of the IRC is called the exposure cycle, which consists of 
one short exposure ($\sim 4.6$~sec) and one long exposure ($\sim 44.4$~sec) 
for the near-infrared channel. 
One pointed observation of spectroscopy contains 8--9 exposure cycles, as described 
in section~\ref{sec:NIR grism Spectroscopic observations with the AKARI/IRC}. 
Short and long exposure data are reduced separately by the toolkit with each set of the calibration parameters. 
Unusual behaviors due to non-linearity or saturation of the detector can be found by comparing 
short- and long-exposure spectra. 
Figure~\ref{fig:spectra-largest} shows the short- and long-exposure spectra of 1~Ceres, 2~Pallas, and 4~Vesta. 
Comparison of the short- and long-exposure spectra suggests that 
the available wavelength range in which both spectra 
match to an accuracy of 10\% of each flux is 
$\lambda < 3.7~\micron$ for 1~Ceres and 
$\lambda < 3.75~\micron$ for 2~Pallas. 
The spectra within these wavelengths can be used for further analyses. 
However, for of 4~Vesta, the short- and long-exposure spectra agree only within $3.2 < \lambda < 3.5~\micron$, 
which 
does not provide necessary information for the present
study. Thus, the 4~Vesta spectra cannot be used for further analyses. 

The spectra of the other asteroids do not display such unusual behavior. 
Specifically, the reduced reflectance spectra (described below) of 1~Ceres and 2~Pallas are consistent with the literature \citep{1997PhDT........10R,2003MPS...38.1383R,2005AA...436.1113V,2006Icar..185..563R,2010Icar..206..327R,2015Natur.528..241D,2015ApJ...804L..13T}.

\subsubsection{Thermal component subtraction}
\label{sec:Thermal Component Subtraction}

The spectrum obtained in the previous section 
essentially consisted of two components in the wavelength range of 2.5--5~$\micron$: 
thermal emission of the asteroid itself ($F_\mathrm{e}$) and reflected sunlight ($F_\mathrm{r}$). 
To obtain the reflectance spectrum, it was necessary to remove the thermal component from the spectrum, which depends on
the distribution of surface temperature at 
the epoch of the observation.
In this work, the Near-Earth Asteroid Thermal Model (NEATM;~\citealt{Harris1998}) was used to 
calculate the thermal flux at longer wavelengths by fitting the spectrum. 
The NEATM is a refinement of the standard thermal model (STM; \citealt{Lebofsky1986}). 
The NEATM solves simultaneously for 
the diameter ($D$) and the geometric albedo ($p_\mathrm{v}$) as well as the beaming parameter ($\eta$) to fit the observed 
(infrared) flux. 
$\eta$ was originally introduced in the STM to allow the model temperature distribution 
to fit the observed enhancement of thermal emission at small solar phase angles (e.g., \citealt{Lebofsky1986}). 
In practice, $\eta$ can be treated as a model parameter that allows a first-order correction for any effect 
that influences the observed surface temperature distribution. 
The geometric information; i.e., the heliocentric distance ($r_\mathrm{h}$), the geocentric distance ($\Delta$), 
and the phase angle ($\alpha$), 
were obtained from JPL Horizons (table~\ref{tab:obslog}). 
The absolute magnitude ($H$) and the slope parameter ($G$) were employed as the visible flux~\citep{Bowell1989}. 
The emissivity was assumed to be constant at $\epsilon=0.9$ throughout the wavelengths considered in this study
(c.f., \citealt{Lebofsky1986}). 
$D$ and $\eta$ were parameterized to fit the thermal flux of the asteroid. 
After subtracting the thermal component fitted by NEATM, the spectrum was divided by the solar spectrum 
based on the corrected Kurucz model~\citep{Berk1999} to obtain the reflectance spectrum of the asteroid. 
The obtained reflectance spectra of each pointed observation are summarized in supplementary data.

\subsubsection{Criterion to reject spectra severely contaminated by thermal emission}
\label{sec:Criterion to reject spectra severely contaminated by thermal emission}
In this research, we aimed to detect small absorption features ($\sim$ 10\%) on 
the spectrum of the reflected sunlight component. The thermal component of the spectra was subtracted as described 
in the previous section. However, it was still necessary to carefully handle the ``contamination'' of 
thermal emission from the asteroid itself. 
It is hard to extract reflectance spectra with a sufficient S/N ratio from spectra with large  thermal emission contribution. 

The grism spectroscopy of the AKARI/IRC covers a wavelength range of 2.5--5 \micron. 
At these wavelengths, 
thermal emission has an equivalent contribution to the total spectrum of a main-belt asteroid to the reflected component. 
As mentioned above, thermal emission was estimated by NEATM; 
that is, thermal emission was 
assumed as a gray body with a fixed emissivity of $\epsilon=0.9$. 
Realistically, the emissivity is not necessarily constant. This ambiguity of emissivity could affect 
the detailed features of the reflectance spectrum. 
We examined the contribution of the thermal component in the total flux density as follows.

Let us consider the relationship between the bi-hemispherical reflectivity ($p_\mathrm{h}$) 
and the hemispherical emissivity ($\varepsilon_\mathrm{h}$) at a certain wavelength. 
As stated by Kirchhoff's law of thermal radiation, $p_\mathrm{h}$ and $\varepsilon_\mathrm{h}$ are complementary as~\citep{Hapke1993} : 
\begin{eqnarray}
\varepsilon_\mathrm{h} &=& 1 - p_\mathrm{h}~. 
\label{eq:e-p}
\end{eqnarray}
Note that both $p_h$ and $\varepsilon_h$ have wavelength dependency; 
$p_\mathrm{h} = p_\mathrm{h} (\lambda)$, and $\varepsilon_\mathrm{h} = \varepsilon_\mathrm{h}(\lambda)$. 
We assume that the spectrum has an absorption feature with a band depth of $s$. 
Here, the band depth is defined as a fraction of the depression from the continuum. 
The bi-directional reflectivity to be observed with the absorption feature ($p_\mathrm{d}$) is given as:
\begin{eqnarray}
p_\mathrm{d} &=& p_\mathrm{d0} (1-s)~, 
\label{eq:p}
\end{eqnarray}
where $p_\mathrm{d0}$ is the continuum of the bi-directional reflectivity. 
In the same way, we assume that the spectrum has an {\it emission} feature with an emission strength of $e$. 
The directional emissivity to be observed with the emission feature ($\varepsilon_\mathrm{d}$) is given as:
\begin{eqnarray}
\varepsilon_\mathrm{d} &=& \varepsilon_\mathrm{d0}(1+e)~,
\label{eq:e}
\end{eqnarray}
where $\varepsilon_\mathrm{d0}$ is the continuum of the directional emissivity. 
It is assumed that an absorption feature appears as downward from the continuum, 
while an emission feature appears as upward. 
Therefore, an absorption is denoted negative ($-s$) in equation~(\ref{eq:p}) and an emission is positive ($+e$) 
in equation~(\ref{eq:e}). 
The bi-hemispherical reflectivity ($p_\mathrm{h}$) and the bi-directional reflectivity ($p_\mathrm{d}$) 
are assumed to have the same 
relation as the Bond albedo to the geometric albedo in the IAU $H$-$G$ function model \citep{Bowell1989} as:
\begin{eqnarray}
p_\mathrm{h} &=& q\,p_\mathrm{d}~, 
\label{eq:p0-pd}
\end{eqnarray}
where $q$ is the phase integral empirically given by the slope parameter ($G$) as:
\begin{eqnarray}
q  &=& 0.290 + 0.684\,G~.
\label{eq:phase integral}
\end{eqnarray}
It should be noted that the applicability of the $H$-$G$ function in the infrared wavelengths 
is not well studied. 
\cite{Lederer2008} reported the photometric properties of S-type asteroid 25143~Itokawa based on ground-based 
UBVRIJHK broadband observational data (0.36-2.15~$\micron$); the difference of the phase integral 
by wavelength is small ($q=0.11 \pm 0.01$ in V-band and $0.13 \pm 0.01$ in K-band). 
There are no other reports to date of the phase integral measured in the near-infrared wavelengths.
Here the wavelength difference of the phase integral of asteroids observed in this study 
is assumed to be small.
Thus equations~(\ref{eq:p0-pd})--(\ref{eq:phase integral}) is applied to all asteroids in our analysis. 


In addition, the angular variation of the directional emissivity ($\varepsilon_\mathrm{d}$) is assumed to be neglected 
(e.g., \citealt{Sobrino1999,Garcia-Santos2012}), thus to be identical to the hemispherical emissivity ($\varepsilon_\mathrm{h}$) as: 
\begin{eqnarray}
\varepsilon_h &\simeq& \varepsilon_d~, 
\end{eqnarray}
and 
\begin{eqnarray}
\varepsilon_\mathrm{h0} &\simeq& \varepsilon_\mathrm{d0}~,
\end{eqnarray}
where $\varepsilon_\mathrm{h0}$ is the continuum of the hemispherical emissivity which is given by Kirchhoff's law as
\begin{eqnarray}
\varepsilon_\mathrm{h0} &=& 1 - p_\mathrm{h0}~, 
\end{eqnarray}
where $p_\mathrm{h0}$ is the continuum of the bi-hemispherical reflectivity and again 
\begin{eqnarray}
p_\mathrm{h0} &=& q\,p_\mathrm{d0}~.
\end{eqnarray}
From these equations, 
\begin{eqnarray}
(1 - q\,p_\mathrm{d0}) (1+e) &=& 1 - q\,p_\mathrm{d0} (1-s)~. \nonumber
\end{eqnarray}
%
%
%
%
%
%
Thus, we have
\begin{eqnarray}
\frac{e}{s} &=& \frac{q\,p_\mathrm{d0}}{1 - q\,p_\mathrm{d0}}~. 
\label{eq:es}
\end{eqnarray}
An asteroid spectrum ($F_\mathrm{tot}$) comprises two components: reflected sunlight ($F_\mathrm{r}$) 
and thermal emission ($F_\mathrm{e}$) as
\begin{eqnarray*}
F_\mathrm{tot} &=& F_\mathrm{r} + F_\mathrm{e}~.
\label{eq:f_tot}
\end{eqnarray*}
Flux densities of the absorption feature ($\Delta F_\mathrm{r}$) and emission feature ($\Delta F_\mathrm{e}$) are written as
\begin{eqnarray*}
\left|\Delta F_\mathrm{r}\right| &=& s F_\mathrm{r}~, \\
\left|\Delta F_\mathrm{e}\right| &=& e F_\mathrm{e}~.  
\end{eqnarray*}
To detect the absorption feature (i.e., measuring $s$) with more than $x \times 100$ \% accuracy, the contribution of 
the emission feature of the thermal component should be suppressed as
\begin{eqnarray*}
\left|\frac{\Delta F_\mathrm{e}}{\Delta F_\mathrm{r}}\right| < x~, 
\end{eqnarray*}
thus
\begin{eqnarray*}
\frac{F_\mathrm{e}}{F_\mathrm{r}} &<& x \frac{1-q\,p_\mathrm{d0}}{q\,p_\mathrm{d0}}~, 
\end{eqnarray*}
or
\begin{eqnarray}
\frac{F_\mathrm{e}}{F_\mathrm{tot}} &<& \frac{1}{\displaystyle\frac{q\,p_\mathrm{d0}}{x(1-q\,p_\mathrm{d0})} + 1}~. 
\label{eq:F_e constraint}
\end{eqnarray}
Here, we consider that $x$ is the same value as the typical uncertainty of the flux density of AKARI spectroscopy.

The geometric infrared albedo ($p_\mathrm{IR}$) is defined as 
the ratio of the brightness at zero phase angle to the brightness of a perfect Lambert disk 
of the equivalent radius at a given infrared wavelength. 
We assume that 
the bi-directional reflectivity of the continuum ($p_\mathrm{d0}$) 
is equal to the infrared albedo ($p_\mathrm{IR}$) at 2.45~$\micron$ which is given as: 
\begin{eqnarray}
p_\mathrm{d0} &=& p_\mathrm{IR} ~\simeq~ p_\mathrm{v} \frac{R_{2.45}}{R_{0.55}}~,  
\label{eq:pir-pv}
\end{eqnarray}
where $p_\mathrm{v}$,  $R_{2.45}$, and  $R_{0.55}$ are 
the geometric visible albedo, the relative reflectance at 2.45~$\micron$, and the relative reflectance at 0.55~$\micron$, 
respectively. 
Note that the second equal sign of equation~(\ref{eq:pir-pv}) comes from the assumption 
that the ratio of the geometric infrared albedo to the geometric visible albedo is as the same as 
the ratio of the relative reflectance at 2.45~$\micron$ and that at 0.55~$\micron$. 
The relative reflectance values at 0.55~$\micron$ and 2.45~$\micron$ are obtained from the vis-NIR spectra 
(spectral data are compiled by \citealt{Hasegawa2017}). 

$\displaystyle\frac{F_\mathrm{e}}{F_\mathrm{tot}}$, which can be calculated by the thermal model 
described in the previous subsection, 
is a function of wavelength. 
Thus, equation~(\ref{eq:F_e constraint}) with $x=0.04$ 
(i.e., 4\% uncertainty; see section~\ref{sec:General trend of the spectra}) is treated as a condition of 
the available wavelength range (upper limit of wavelength, $\lambda_\mathrm{trunc}$) 
to extract the reflectance spectrum with sufficient accuracy.
$\lambda_\mathrm{trunc}$ values for each observed target are listed in table~\ref{tab:thermal_contamination}. 
It is natural that 4015~Wilson-Harringtion, which is the near-Earth object, has a high surface temperature; thus, the thermal component fully occupies the wavelength coverage of the spectroscopy ($\lambda_\mathrm{trunc}=3.0~\micron$). 
We cannot retrieve any valid reflectance spectra from this object; instead, the spectral data of this object were utilized for studying the thermal properties of the asteroid (e.g., \citealt{Bach2017}). 

\subsubsection{3-$\micron$-band depth measurements}
\label{sec:3um-band depth measurements}
The measurement method for band depth described in~\cite{2012Icar..219..641T} 
cannot be applied to our data set because our spectra only cover wavelengths 
longer than 2.5~$\micron$, 
which is the starting wavelength for absorption features associated with 
hydrated minerals. 
Instead, the band depth ($\mathcal{D}$) was measured as: 
\begin{enumerate}
\item before measuring the band depth, two or three reflectance spectra of each asteroid were averaged,  
\item the continuum ($R_c$) was defined as the connection of two local peaks of the reflectance spectrum 
across the peak wavelength of the absorption feature, 
\item the band depth ($\mathcal{D}$) was defined as (see figure~\ref{fig:data-example})
\begin{eqnarray}
\mathcal{D} &=& \max \left(\frac{R_c - R_\lambda}{R_c}\right)\ ,
\end{eqnarray} 
where  $R_\lambda$ is the reflectance spectrum (c.f., \citealt{Clark1984}). 
\end{enumerate}
The peak wavelength was defined as the wavelength with the maximum value of the band depth. 
To increase the S/N, two or three spectra obtained were averaged in step 1. 
Thus, the spectral variability due to heterogeneities of the surface material or any other reason was smoothed out. 
After a visual inspection of our results, 
obvious features appeared at around 2.7~$\micron$ and 3.1~$\micron$ in the spectra. Thus, two band depths,
$\mathcal{D}_{2.7}$ and $\mathcal{D}_{3.1}$, are treated as absorption features in this study and 
small features in other wavelengths are not considered.

\section{Results}
\subsection{General trend of spectra}
\label{sec:General trend of the spectra} 
\begin{figure}
\begin{center}
\includegraphics[height=110mm]{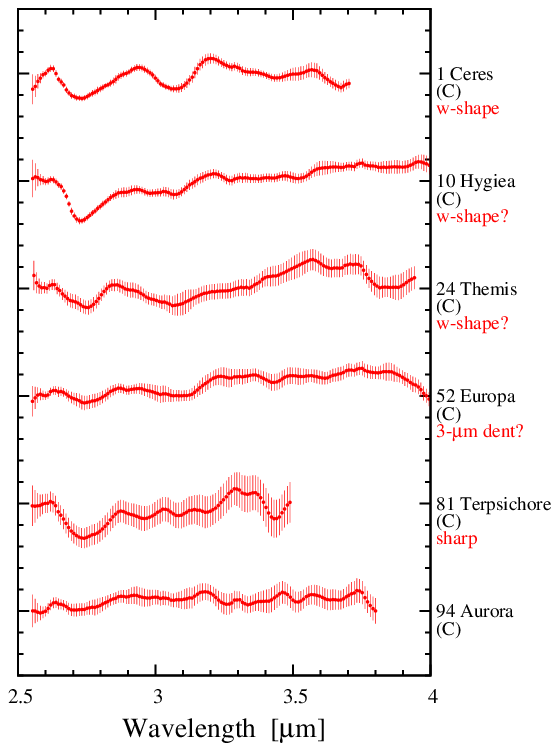}\includegraphics[height=110mm]{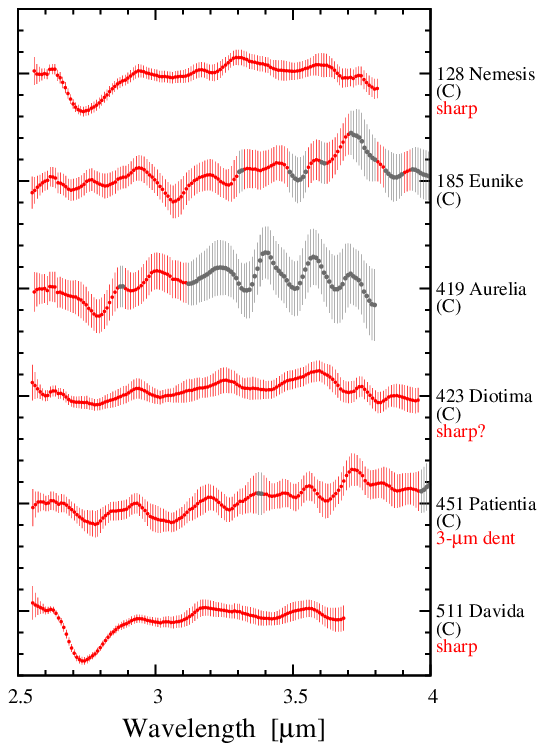}\\
\includegraphics[height=110mm]{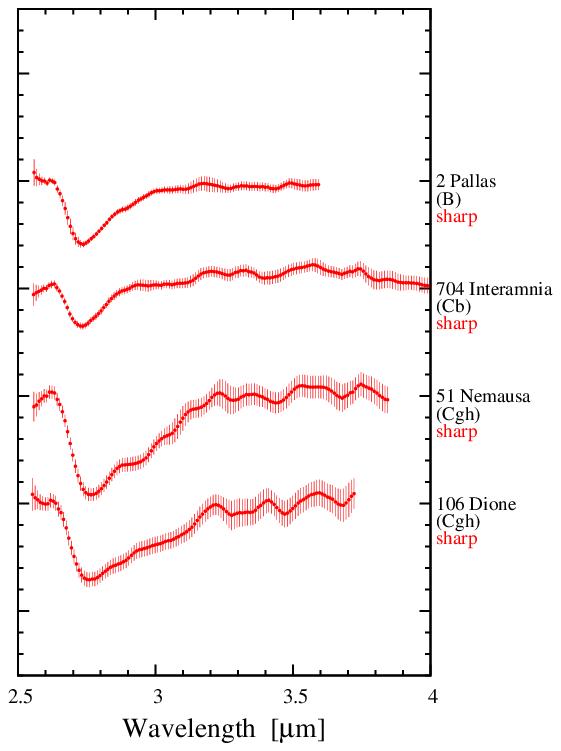}\includegraphics[height=110mm]{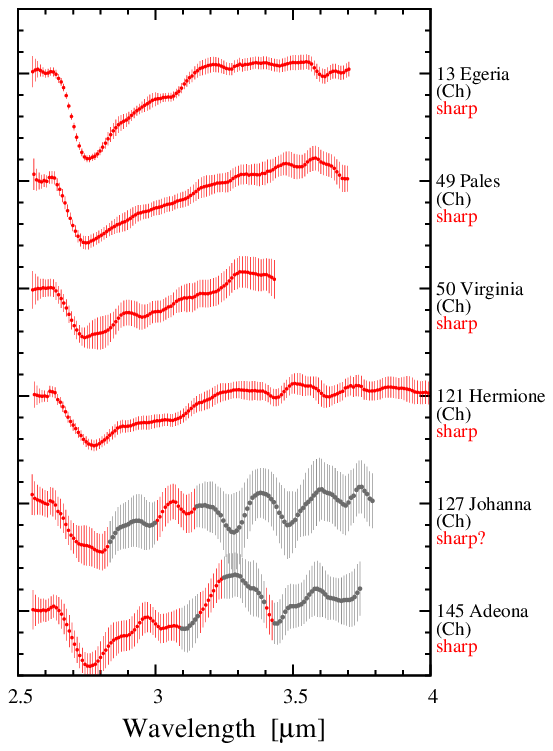}\\
\end{center}
\caption{Average reflectance spectra of C-complex asteroids. The Bus-DeMeo taxonomy is indicated in parentheses (see table~\ref{tab:object_tax}). 
The gray dots indicate the unreliable wavelength region due to large uncertainties ($> 10$\%). 
The 3-$\micron$ band shape is given in red text (see table~\ref{tab:band_depth}).}\label{fig:spec-refl-sum:Ctype}
\end{figure}

\begin{figure}
\begin{center}
\includegraphics[height=110mm]{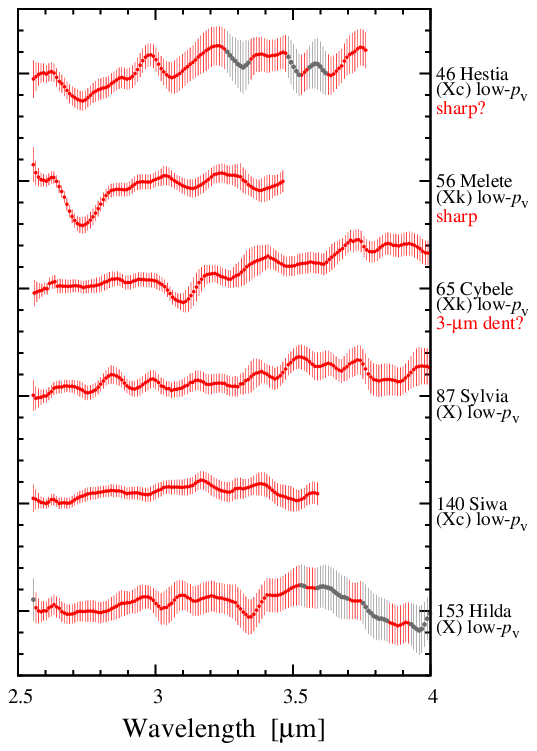}\includegraphics[height=110mm]{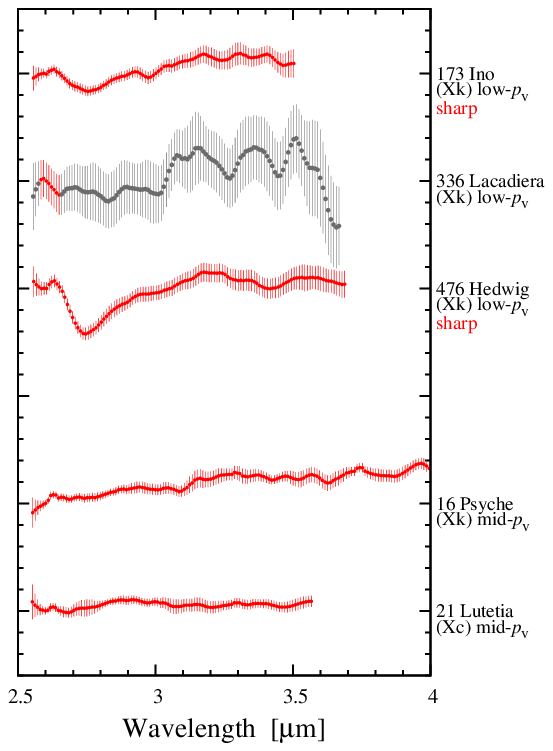}\\
\includegraphics[height=110mm]{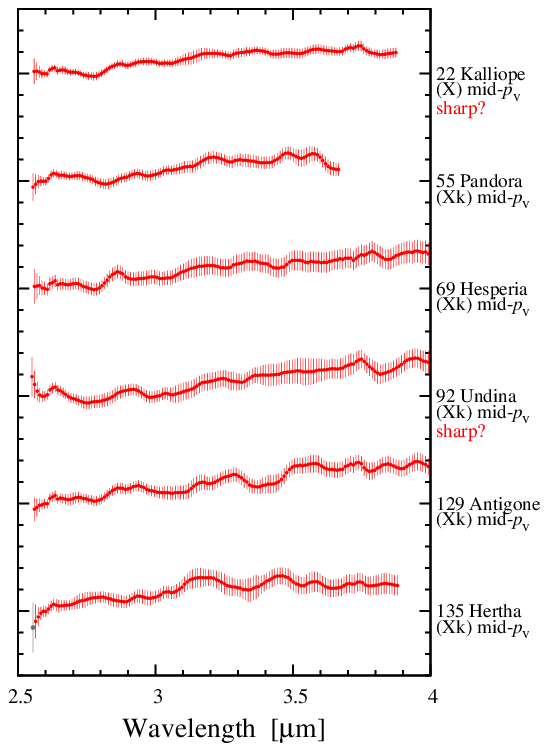}\includegraphics[height=110mm]{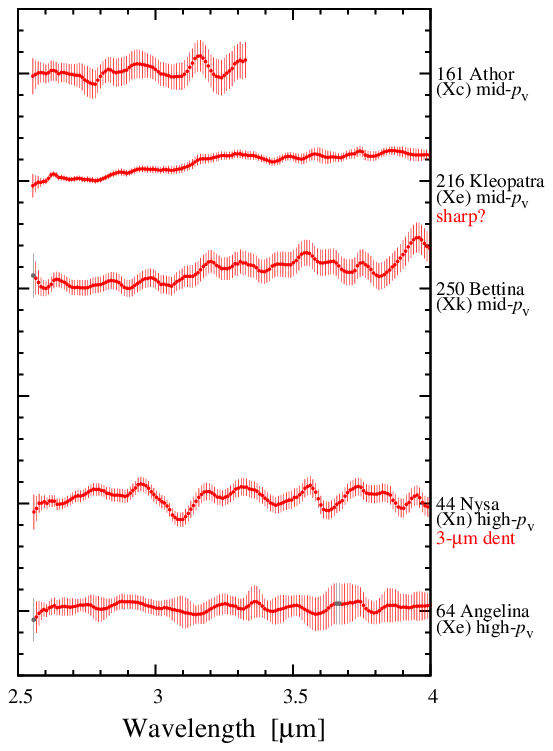}\\
\end{center}
\caption{Same as Fig.~\ref{fig:spec-refl-sum:Ctype} but for X-complex asteroids. 
The objects are labeled by albedo value: 
low-$p_\mathrm{v}$ ($p_\mathrm{v} < 0.1$), 
mid-$p_\mathrm{v}$ ($0.1 < p_\mathrm{v} < 0.3$), 
and 
high-$p_\mathrm{v}$ ($p_\mathrm{v} > 0.3$). 
}\label{fig:spec-refl-sum:Xtype}
\end{figure}

\begin{figure}
\begin{center}
\includegraphics[height=110mm]{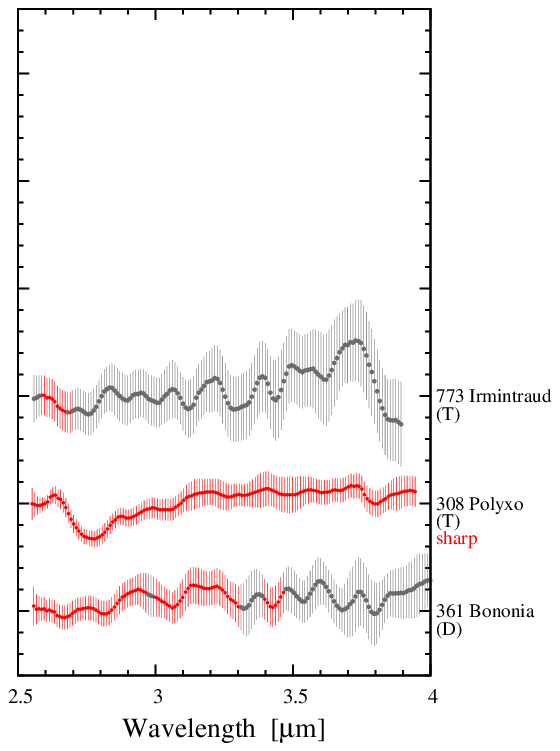}
\end{center}
\caption{Same as Fig.~\ref{fig:spec-refl-sum:Ctype} but for D-complex asteroids.}\label{fig:spec-refl-sum:Dtype}
\end{figure}

\begin{figure}
\begin{center}
\includegraphics[height=110mm]{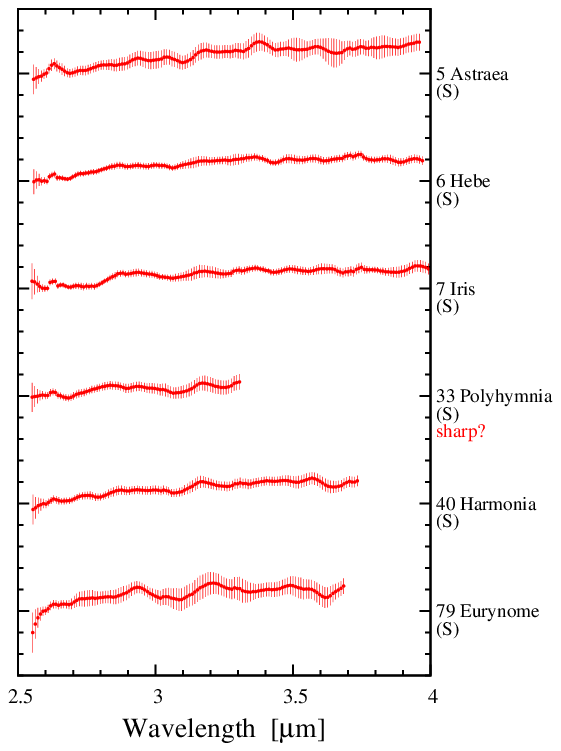}\includegraphics[height=110mm]{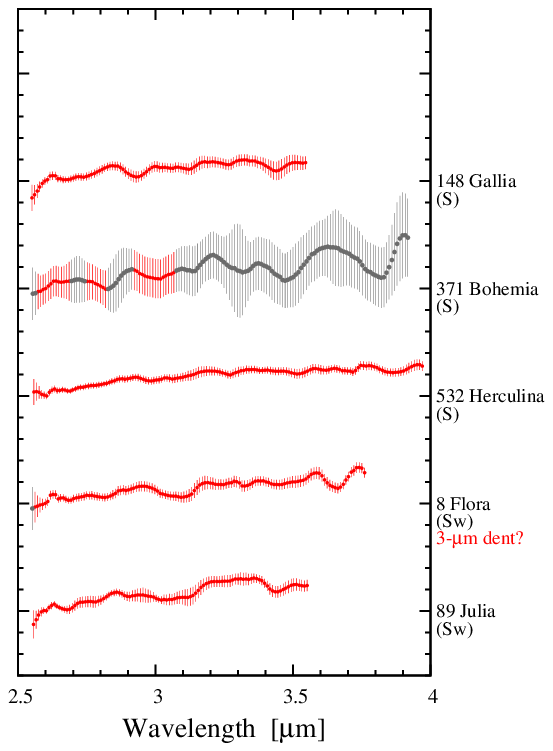}\\
\includegraphics[height=110mm]{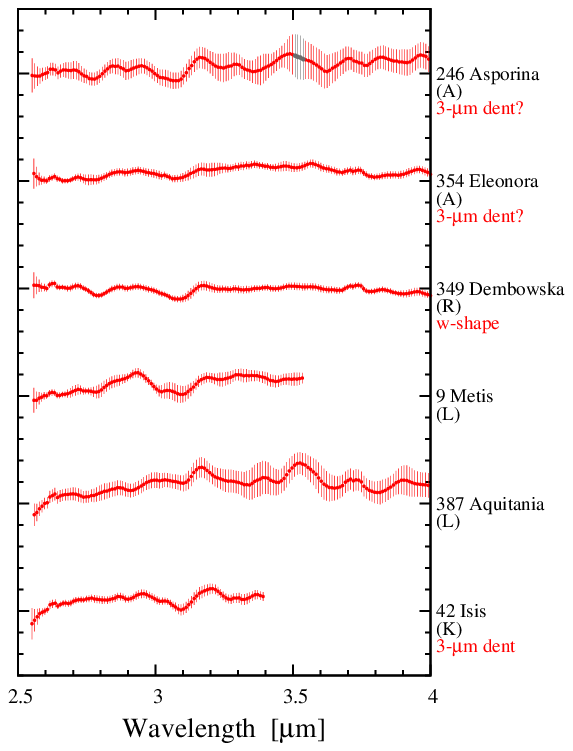}
\end{center}
\caption{Same as Fig.~\ref{fig:spec-refl-sum:Ctype} but for S-complex asteroids.}\label{fig:spec-refl-sum:Stype}
\end{figure}

\begin{figure}
\begin{center}
\includegraphics[width=140mm]{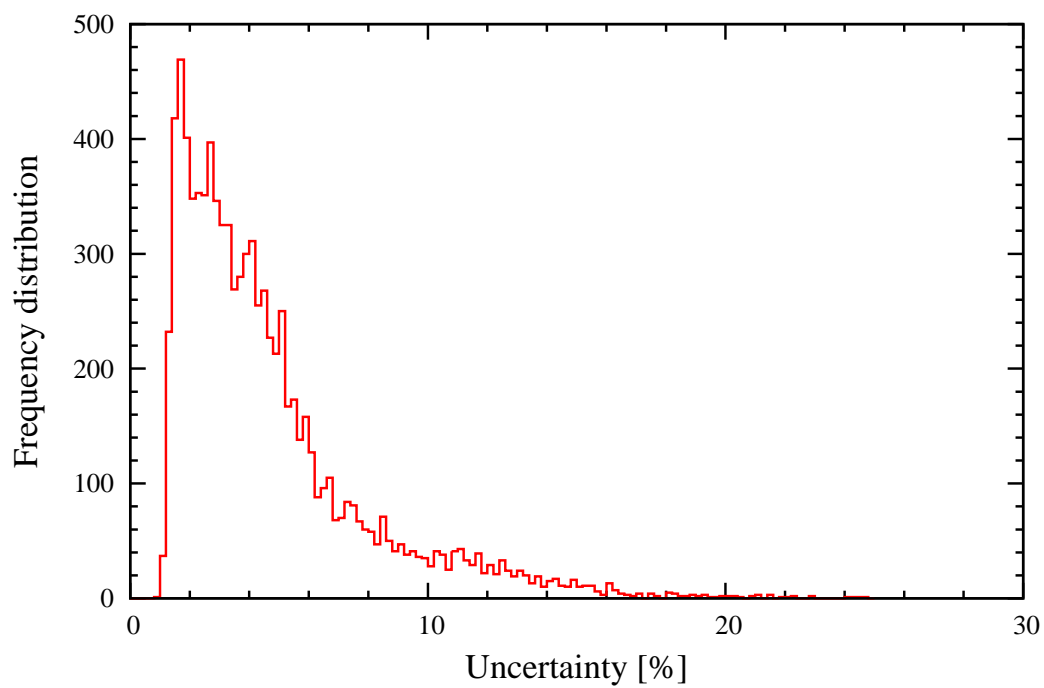}
\end{center}
\caption{
Distribution of the uncertainties in the data points of all the reflectance spectra of 64 asteroids. 
The spectra within $< \lambda_\mathrm{trunc}$ are used for this histogram. 
}\label{fig:uncertainty distribution}
\end{figure}

The obtained reflectance spectra are summarized in figures~\ref{fig:spec-refl-sum:Ctype}-\ref{fig:spec-refl-sum:Stype}. 
Figure~\ref{fig:uncertainty distribution} shows 
the distribution of uncertainties of all the data points in the reflectance spectra of 64 asteroids. 
Most of the data points in the reflectance spectra have uncertainties smaller than 10\%
(the median value is 3.8\%). 
The uncertainties in the reflectance spectra are caused by 
the uncertainty in the absolute flux calibration ($\sim$5\%), 
the ambiguity of the wavelength zero point on the reference image ($\sim$8\%), 
and 
the fluctuation of the sky background ($\sim$3\%), 
which are described in section~\ref{sec:Extraction of one-dimensional spectra}. 
Uncertainty in the thermal model calculation described in section~\ref{sec:Thermal Component Subtraction} 
is not taken into account in the reflectance spectra; 
this uncertainty contributes to the spectra at $\lambda_\mathrm{trunc}$ or longer wavelengths. 

As described above, 
the observed spectrum of 4~Vesta (V-type) is saturated in the 3-$\micron$ band (section~\ref{sec:saturation}), and 
the thermal contamination of 4015~Wilson-Harringtion (B-type) is not fully corrected 
(section~\ref{sec:Criterion to reject spectra severely contaminated by thermal emission}). 
Therefore, the spectra of these two objects cannot be used for further analysis. 
In total, the reflectance spectra of 64 objects are available in this study. 

Diagnostic spectral features appear at wavelengths
from 2.6~$\micron$ to 2.9~$\micron$ or longer with a band center of $\sim$ 2.75~$\micron$
(hereafter the 2.7-$\micron$ band), and from 2.8--2.9~$\micron$ to 3.2~$\micron$ or longer with a band center of $\sim$ 3.05~$\micron$ (hereafter the 3.1-$\micron$ band). 
Our results confirm that no significant spectral feature with broad peak centered 
at 2.95~$\micron$ is found in the asteroid spectra, while such a feature sometimes 
appears in the spectra of meteorites, which is attributed to adsorbed water 
\citep{Beck2010, Rivkin2015AIV}. 
There are other absorption features in the 3-$\micron$ band; for example, those associated with 
organic materials at 3.4--3.6~$\micron$ (e.g., \citealt{DeSanctis2017}), which is another aspect 
of interest to this study. 

Most C-complex asteroids have clear absorption features 
in the 2.7-$\micron$ band, as well as some in the 3.1-$\micron$ band. 
On the contrary, most S-complex asteroids show no significant features at these wavelengths. 
Some X-complex asteroids have 
absorption features in the 2.7-$\micron$ band, but among D-complex asteroids, only one asteroid in our data shows 
an absorption feature in the 2.7-$\micron$ band. 

\begin{figure}
\begin{center}
\includegraphics[width=140mm]{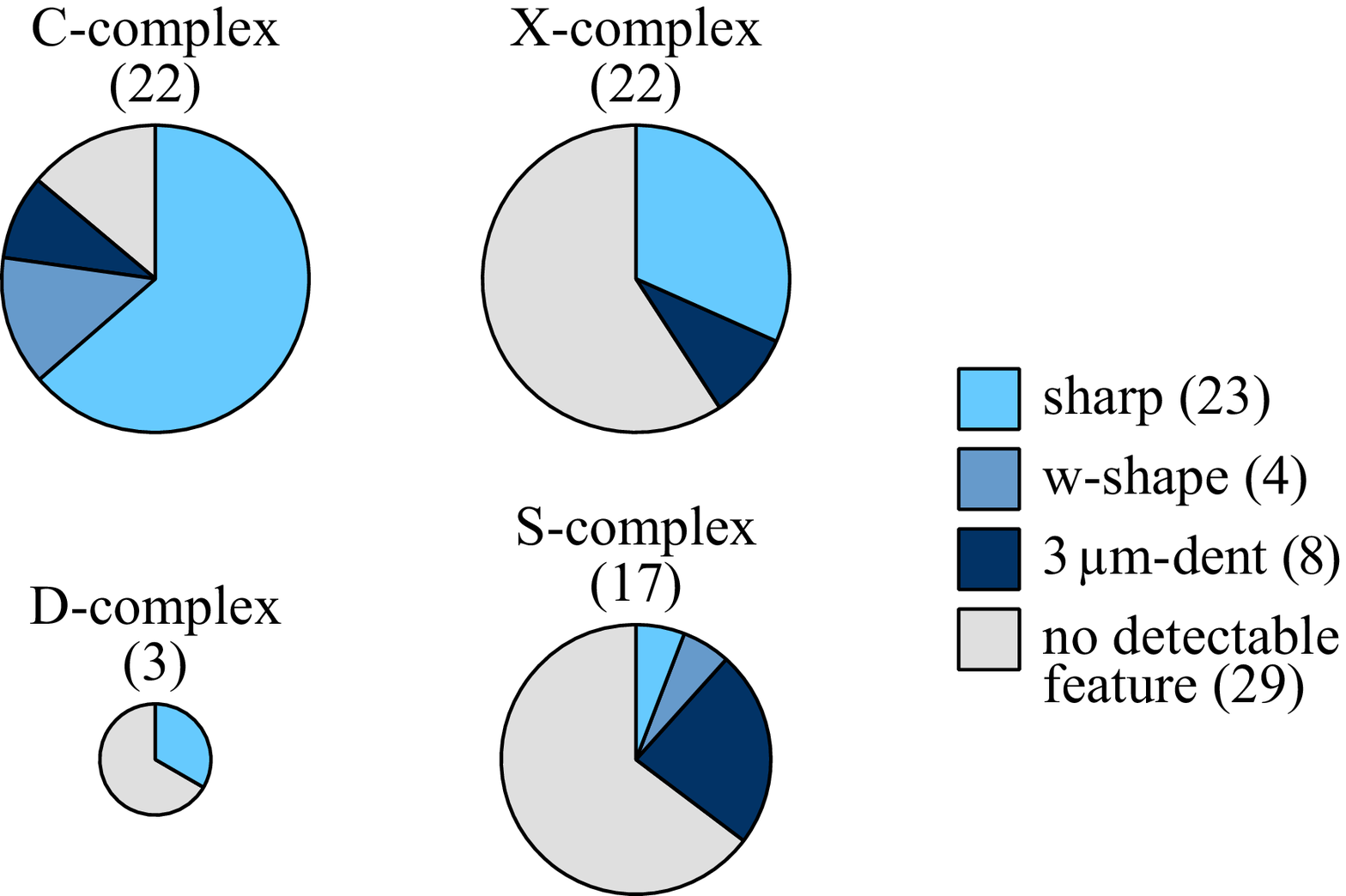}
\end{center}
\caption{
Number of the 3-$\micron$ band shape classified in this study. 
}\label{fig:pie-chart}
\end{figure}

\begin{figure}
\begin{center}
\includegraphics[width=140mm]{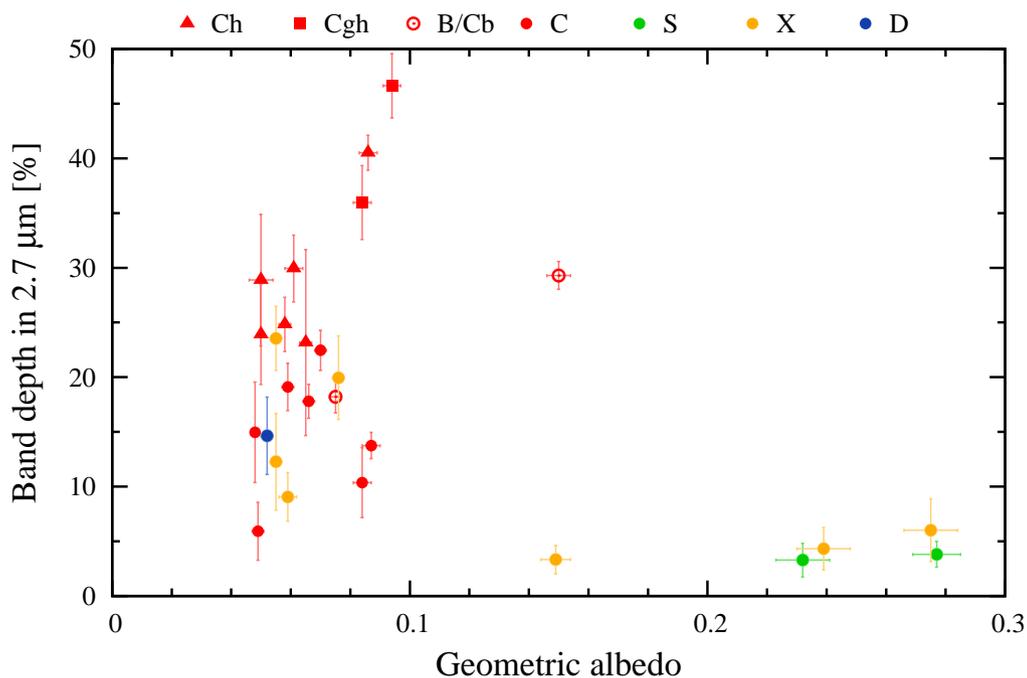}
\end{center}
\caption{
Distribution of band depths in the 2.7-$\micron$ band of reflectance spectra (S/N $> 2$) 
against geometric albedo. 
The red, green, yellow, and blue symbols denote C-, S-, X-, and D-complex asteroids, respectively. 
C-complex asteroids are subdivided into rectangles, triangles, open circles, and filled circles, as 
Cgh-, Ch-, B or Cb, and other C-type, respectively. 
The correlation coefficient is -0.46. 
}\label{fig:strength-albedo} 
\end{figure}

\begin{figure}
\begin{center}
\includegraphics[width=140mm]{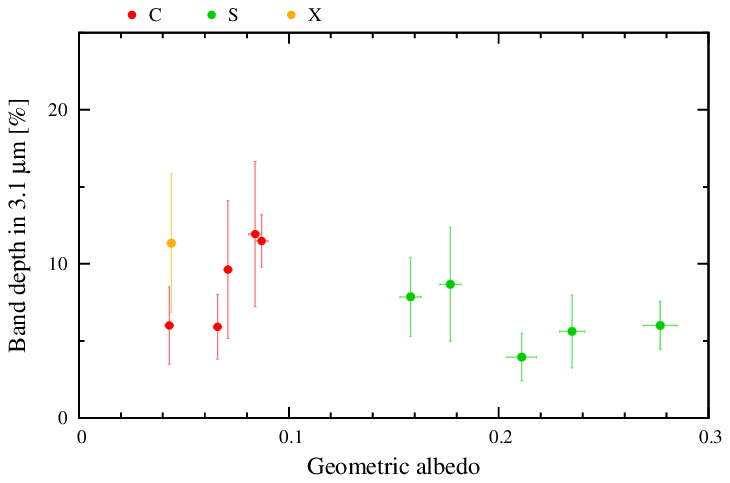}
\end{center}
\caption{
Same as figure~\ref{fig:strength-albedo} but in the 3.1-$\micron$ band. 
The correlation coefficient is 0.11. 
}\label{fig:strength-albedo:3.1um} 
\end{figure}

\begin{figure}
\begin{center}
\includegraphics[width=140mm]{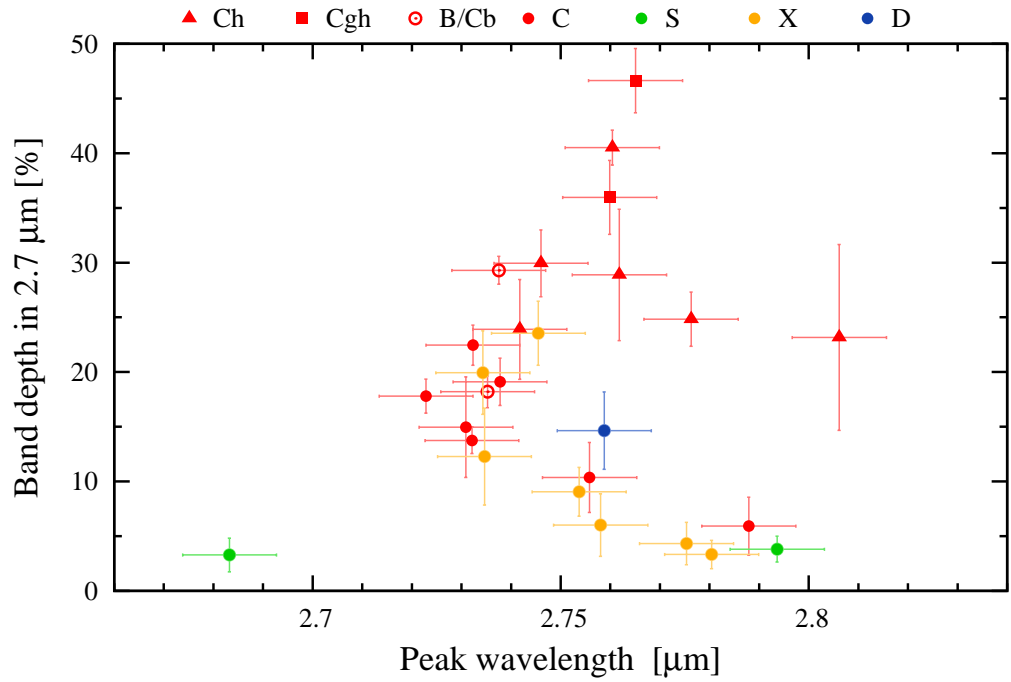}
\end{center}
\caption{
Same as figure~\ref{fig:strength-albedo} but against the peak wavelength of band depth in the 2.7-$\micron$ band.
The correlation coefficient is 0.02.
}\label{fig:strength-peakwavelength} 
\end{figure}

\begin{figure}
\begin{center}
\includegraphics[width=140mm]{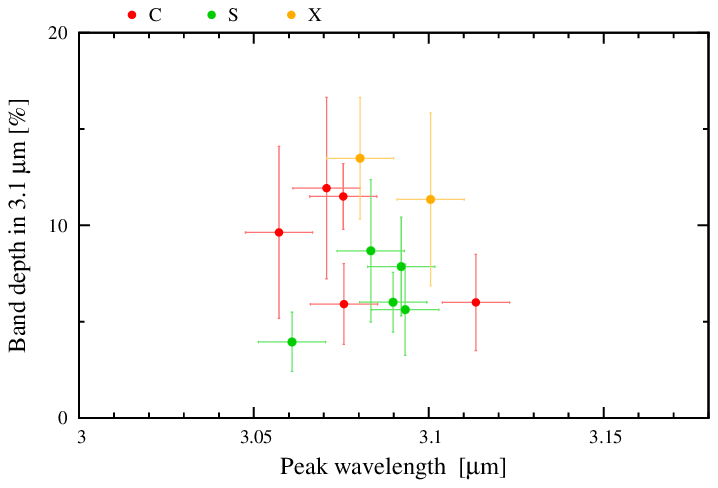}
\end{center}
\caption{
Same as figure~\ref{fig:strength-peakwavelength} but in the 3.1-$\micron$ band. 
The correlation coefficient is -0.12.
}\label{fig:strength-peakwavelength:3.1um} 
\end{figure}

\begin{figure}
\begin{center}
\includegraphics[width=140mm]{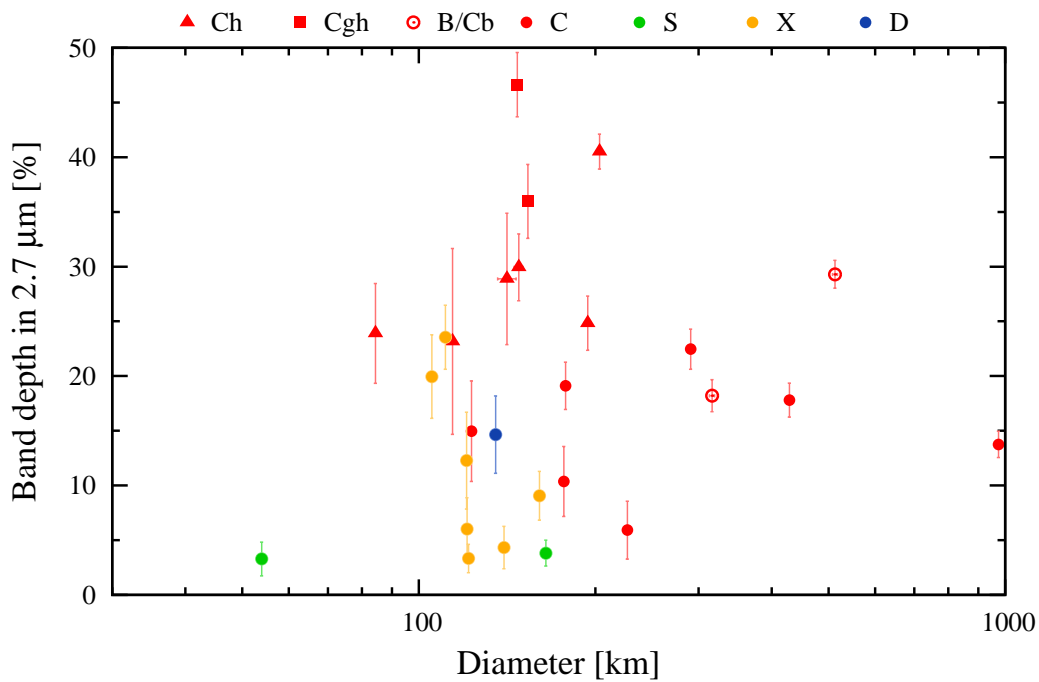}
\end{center}
\caption{
Same as figure~\ref{fig:strength-albedo} in the 2.7-$\micron$ band but against the asteroid diameter.
The correlation coefficient is 0.03.
}\label{fig:strength-size} 
\end{figure}

\begin{figure}
\begin{center}
\includegraphics[width=140mm]{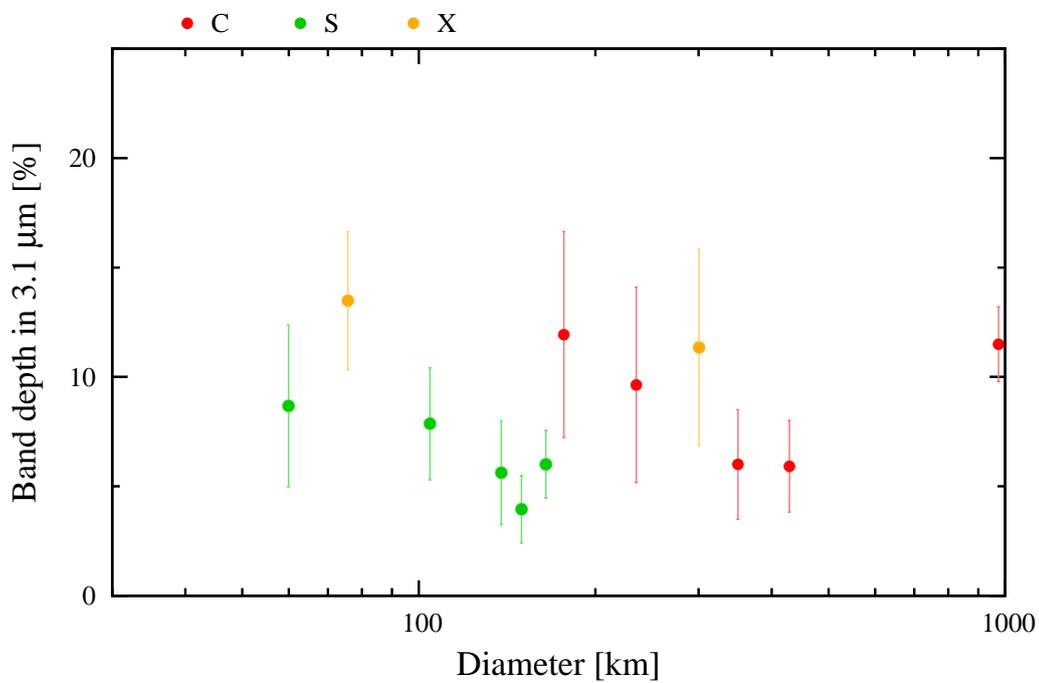}
\end{center}
\caption{
Same as figure~\ref{fig:strength-size} but in the 3.1-$\micron$ band. 
The correlation coefficient is 0.18. 
}\label{fig:strength-size:3.1um} 
\end{figure}

\begin{figure}
\begin{center}
\includegraphics[width=140mm]{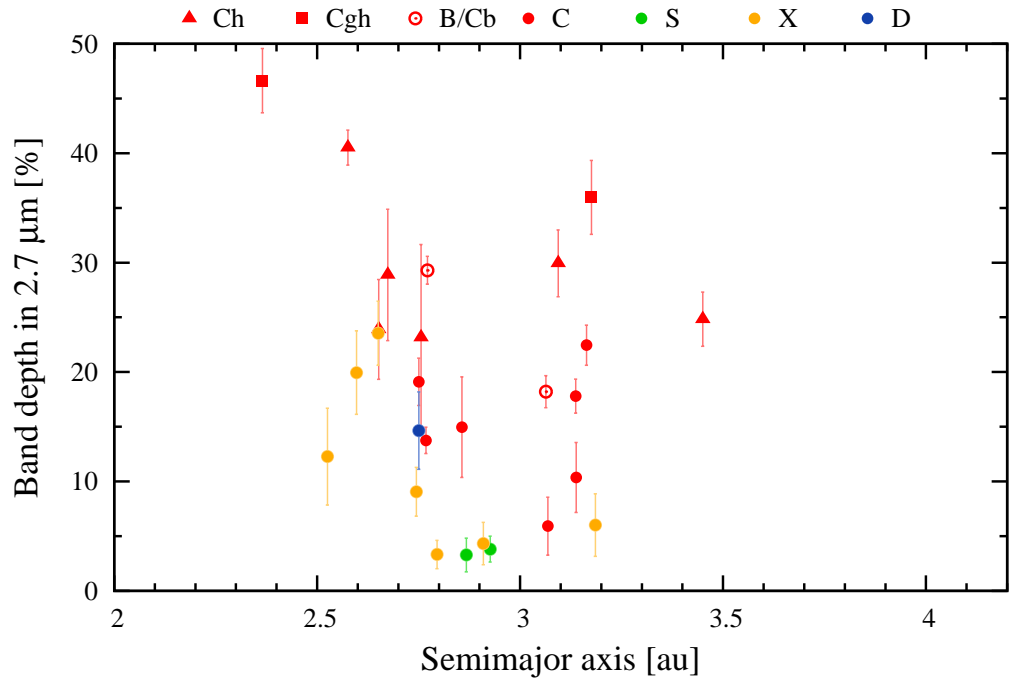}
\end{center}
\caption{
Same as figure~\ref{fig:strength-albedo} in the 2.7-$\micron$ band but against the semimajor axis of asteroids. 
The correlation coefficient is -0.25.
}\label{fig:strength-semj} 
\end{figure}

\begin{figure}
\begin{center}
\includegraphics[width=140mm]{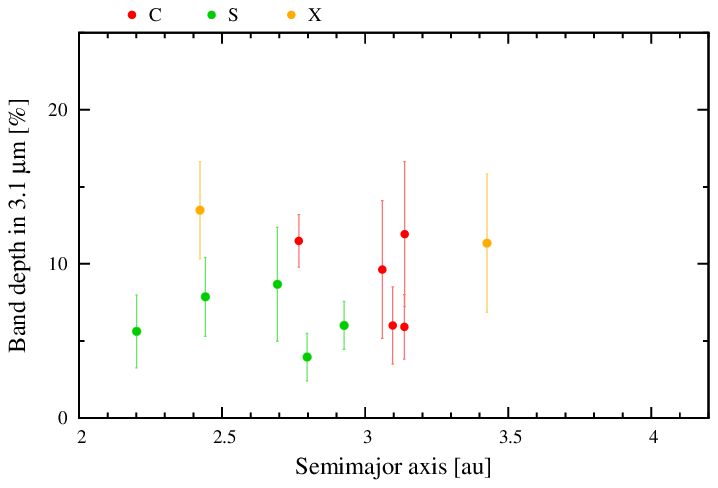}
\end{center}
\caption{
Same as figure~\ref{fig:strength-semj} but in the 3.1-$\micron$ band.
The correlation coefficient is 0.11. 
}\label{fig:strength-semj:3.1um} 
\end{figure}

\begin{figure}
\begin{center}
\includegraphics[width=140mm]{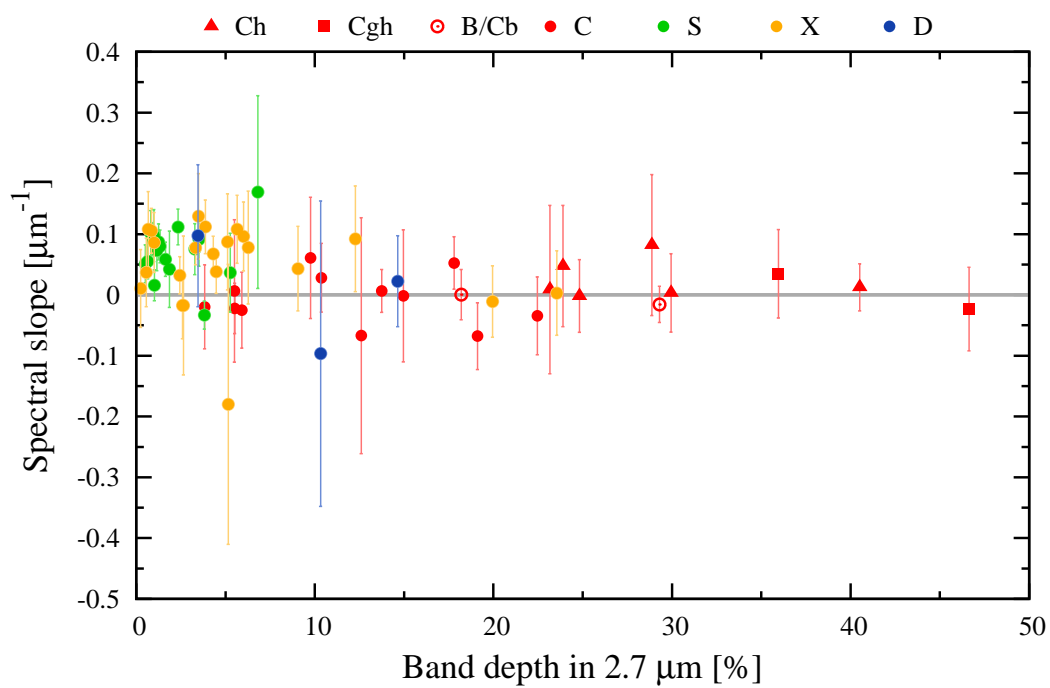}
\end{center}
\caption{
Distribution of the spectral slope 
measured from 2.6~$\micron$ to $\lambda_\mathrm{trunc}$ in the reflectance spectra 
against the band depth in the 2.7-$\micron$ band. 
The solid gray line indicates a horizontal line of zero slope ($\mathcal{S}=0$). 
To clarify the plot, the error bars of band depth (in the horizontal direction) are omitted. 
The correlation coefficient is -0.33. 
}\label{fig:slope-strength} 
\end{figure}

The band depths and the peak wavelengths of spectra in the 2.7- and 3.1-$\micron$ bands 
for each asteroid are summarized in table~\ref{tab:band_depth}. 
The 3-$\micron$ band shape in this table 
is determined by a combination of the spectral features 
in 2.7~$\micron$ ($\mathcal{D}_{2.7}$) and 3.1~$\micron$ ($\mathcal{D}_{3.1}$) bands as: 
\begin{itemize}
\item {\bf sharp} spectral features with only $\mathcal{D}_{2.7}$,  
\item {\bf w-shape} spectral features
\footnote{Note that ``w''-shape in this work is determined by the appearance of its spectrum, and is not intended to relate to ``W''-class (wet M-class) asteroids~\citep{1995Icar..117...90R,2000Icar..145..351R}.} 
with both $\mathcal{D}_{2.7}$ and $\mathcal{D}_{3.1}$,  
\item {\bf 3-}{\boldmath $\micron$}~{\bf dent} features with only $\mathcal{D}_{3.1}$,  
\item unclassified. 
\end{itemize}
Here, detection of the band depth is identified by an S/N signal $> 2$. 
The number of classified objects is 
summarized in figure ~\ref{fig:pie-chart}; 
sharp ($\times 23$), w-shape ($\times 4$), and 3-$\micron$ dent ($\times 8$) out of 64 asteroids. 
The relationships between band depth 
and other physical parameters are shown in 
figures~\ref{fig:strength-albedo}--\ref{fig:slope-strength}. 

Figure~\ref{fig:strength-albedo} shows the relationship between the band depth in the 2.7-$\micron$ band 
and the geometric albedo. From this figure, a general trend of 
absorption in the 2.7-$\micron$ band is observed, whereby 
most C-complex asteroids (17 out of 22), three X-complex asteroids, and one D-complex asteroid have significant absorption features, 
which is typically associated with hydrated minerals. 
All these objects 
have low albedo ($p_\mathrm{v} < 0.1$), 
except for 2~Pallas ($p_\mathrm{v} = 0.15$; B-type). 
However, removing the top three absorption objects (51~Nemausa: 47\%, 13~Egeria: 41\%, and 106~Dione: 36\%), there is no clear correlation found.	
Figure~\ref{fig:strength-albedo:3.1um} reveals the relationship in the 3.1-$\micron$ band, in which 
no clear correlation is observed. 

Figures~\ref{fig:strength-peakwavelength} and \ref{fig:strength-peakwavelength:3.1um} 
show the distribution of peak absorption wavelengths for the 2.7- and 3.1-$\micron$ bands, respectively.
The mean value of peak wavelength in the 2.7-$\micron$ band of 27 objects with S/N $> 2$ 
is 
$2.75 \pm 0.03~\micron$, and that in the 3.1-$\micron$ band of 12 objects is 
$3.08 \pm 0.02~\micron$. 
The uncertainty of the peak wavelength is estimated by a 1-pixel resolution of approximately 0.01~$\micron$. 

Figure~\ref{fig:strength-size} shows the relationship with the size of objects. 
Based on the size information from \cite{Usui2011}, the largest 12 objects are analysed in this study, 
except for the following two objects: 
31~Euphrosyne (276~km; Cb-type, \citealt{Bus2002})
and 
15~Eunomia (256~km; K-type, \citealt{DeMeo2009}). 
\cite{2012Icar..219..641T} reported that 31~Euphrosyne is classified as Europa-like with a
13.63\% band depth in the 3.00-$\micron$ band. 
These largest objects, with a diameter $>$ 250~km, have a band depth in the 2.7-$\micron$ band of moderate to weak 
($30 > \mathcal{D}_{2.7} > 10\%$), while some 150--250~km objects 
have greater band depths ($\mathcal{D}_{2.7} > 30\%$). 
On the other hand, no clear trend is observed in the distribution of the 3.1-$\micron$ band
in figure~\ref{fig:strength-size:3.1um}. 

Figures~\ref{fig:strength-semj} and \ref{fig:strength-semj:3.1um} show the relationship with the semimajor axis of objects. 
There are no clear trends in this distribution because the scatter is too large and the data suffered from significant observational bias. 

Figure~\ref{fig:slope-strength} displays the spectral slope measured in the 3-$\micron$ band. 
In this study, the spectral slope ($\mathcal{S}$) is defined as the slope from 2.6~$\micron$ to $\lambda_\mathrm{trunc}$ 
described in 
section~\ref{sec:Criterion to reject spectra severely contaminated by thermal emission}. 
The mean value of the slope of the asteroids which have an absorption feature in the 2.7-$\micron$ band 
($\mathcal{D}_{2.7} > 10\%$) is $\overline{\mathcal{S}}=0.004 \pm 0.033$,
which is almost flat. 
In contrast, that of the asteroids which do not have a feature ($\mathcal{D}_{2.7} < 10\%$) is 
a positive slope ($\overline{\mathcal{S}}=0.058 \pm 0.044$), which is 
likely to be connected with the slope of vis-NIR spectra. 

\subsection{Individual objects}
Brief descriptions of 64 individual asteroids are given in this section. 

\subsubsection{C-complex asteroids}

Most C-complex asteroids (17 out of 22),
especially all Ch-, Cgh-, B-, and Cb-type asteroids, 
have obvious absorption features ($> 18\%$) at around 2.75~$\micron$. 
Among C-complex asteroids, four types of spectral feature are observed: 
sharp, w-shape, 3-$\micron$ dent, and no clear significant feature. 
Moreover, in the sharp group, two subgroups exist:
spectra with relatively large band depth ($\gtrsim$ 24\%), whose features extend toward 3~$\micron$ or longer wavelengths, and 
spectra with moderate band depth ($\lesssim$ 24\%),
whose features end at around $\sim$ 2.9~$\micron$. 
The former group either does not exhibit the 3.1-$\micron$ band feature 
or it is obscured by the strong 2.7-$\micron$ band feature. 
The latter may have other features longer than 3~$\micron$ (weak 3-$\micron$ dent). 
The former comprises Ch- and Cgh-types, while the latter belongs to other C-complex asteroids. 

\begin{itemize}
\item C-type asteroids
\begin{description}
\item{1 Ceres}\\
Ceres, classified as a dwarf planet, is the largest object in the main-belt region. 
This object is classified as a C-type asteroid and was the first object observed in the 3-$\micron$ band; i.e., \cite{1978MNRAS.182P..17L} clearly detected 
the 3-$\micron$ depth on its spectrum, \cite{King1992} discussed the existence of ammoniated phyllosilicate on Ceres, and \cite{2015ApJ...804L..13T} discussed phase angle-induced spectral effects on the 3-$\micron$ absorption band.
Recently, Ceres has been comprehensively studied using in-situ observations of 
the spacecraft Dawn~\citep{Russell2004, Russell2011}; 
an absorption band centered near 3.1~$\micron$ has been attributed to 
ammoniated phyllosilicates widespread across its surface~\citep{2015Natur.528..241D}. 

The characteristics of the spectrum taken with AKARI are essentially consistent with those of previous researches: 
the band centers located at 2.73~$\micron$ (2.72~$\micron$, \citealt{2015Natur.528..241D}) and 
3.08~$\micron$ (3.06--3.07~$\micron$, \citealt{2015ApJ...804L..13T}). 
In this study, this spectral feature is 
categorized as the w-shape. 
There is a bumpy structure at around 3.2~$\micron$ of uncertain origin, 
which is the similar feature found in \cite{2015ApJ...804L..13T} observed at the same phase angle of $\alpha \sim 22^{\circ}$. 

\item{10 Hygiea}\\
Hygiea is a C-type asteroid, the fourth largest object (diameter of 430 km) in the main-belt region 
and is the largest member of its own family~\citep{Carruba2013}. 
Hygiea is a slow rotator, with one revolution lasting 27.63~hr~\citep{Warner2009}. 
Hygiea has a fairly homogeneous surface~\citep{Mothe-Diniz2001}. 
\cite{2012Icar..219..641T} classified the 3-$\micron$ band spectral feature of this object into the Ceres-like group. 
Our AKARI results also show that this object has two significant features at 2.72 and 3.08~$\micron$; thus, this 
spectral feature is also categorized as the w-shape.

\footnotetext[4]{Nesvorny, D., 2015, NASA Planetary Data System, EAR-A-VARGBDET-5-NESVORNYFAM-V3.0}
\addtocounter{footnote}{1}

\item{24 Themis}\\
Themis is a C-type asteroid. 
It is the largest member of its own family (e.g., D. Nesvorny 2015\footnotemark[4]). 
\cite{Fornasier1999} reported that this object has a signature of aqueous alteration based on ground-based observations 
at visible wavelengths. 
\cite{2012Icar..219..641T} classified the 3-$\micron$ band spectral feature of this object into the rounded group. 
The rounded shape has been previously identified by \cite{2010Natur.464.1320C}. 
\cite{2010Natur.464.1322R} found that the spectrum of Themis is matched 
by a spectral model of a mixture of ice-coated pyroxene grains and amorphous carbon, 
suggesting that the surfaces of these asteroids with rounded features include 
very fine water frost in the form of grain coatings. 
Our AKARI results show that its spectrum has two absorption features: at around 2.76~$\micron$, which is associated with 
hydrated minerals, and at around 3.07~$\micron$, a wide feature related to water ice; 
thus, this 
spectral feature is categorized as the w-shape. The 3.1-$\micron$ feature 
of this object, which is thought to be associated with water ice, is broader than that of 1~Ceres or 10~Hygiea. 

\item{52 Europa}\\
Europa is one of the largest C-type asteroids (diameter of 350 km). 
This object has a detailed 3-D shape model based on Adaptive Optics (AO) imaging observations made with the Keck telescope~\citep{Merline2013}. 
D. Nesvorny (2015)\footnotemark[4] reported that this belongs to the Hygiea family. 
\cite{2012Icar..219..641T} classified the 3-$\micron$ band spectral feature of this object into the Europa-like group 
using its own name. 
Our results show a small 3.1-$\micron$ feature as well as a weaker feature at 2.7~$\micron$, which 
is below the detection limit (S/N$< 2$). 

\item{81 Terpsichore}\\
Terpsichore is a C-type asteroid and the largest member of 
the asteroid family bearing its name (D. Nesvorny 2015\footnotemark[4]). 
The albedo is 0.048 according to AKARI~\citep{Usui2011} or 0.034 according to WISE~\citep{Masiero2012}, which is relatively dark as C-complex asteroids. 
There are no reports to date of observations in the 3-$\micron$ band. 
Our AKARI results show the 2.7-$\micron$ feature, which is classified into the sharp group. 

\item{94 Aurora}\\
Follow-up observations at 1--2.5~$\micron$ were performed by~\cite{Hasegawa2017}, and this object was classified 
as C-type according to the Bus-DeMeo taxonomy. 
No features have been observed yet in the 3-$\micron$ band; our results also show no significant feature in the 3-$\micron$ band. 

\item{128 Nemesis}\\
Nemesis is a C-type asteroid and the largest member of the asteroid family bearing its name (D. Nesvorny 2015\footnotemark[4]).
Nemesis is a slow rotator, taking 77.81~hr for one revolution~\citep{Warner2009}. 
Heterogeneous surface properties were reported for this object by~\cite{Scaltriti1979}.
Our AKARI results show a significant feature in 2.74~$\micron$, which is classified into the sharp group. 

\item{185 Eunike}\\
Eunike is a C-type asteroid and a slow rotator, taking 21.812~hr for one revolution~\citep{Warner2009}. 
Albedo variegation of this object was reported by \cite{Pilcher2014}. 
\cite{Fornasier1999} reported that this object does not exhibit an aqueous alteration signature according to ground-based observations at visible wavelengths. 
\cite{2003MPS...38.1383R} reported an absorption feature of 5.3\% towards 2.7~$\micron$, 
but our results show no feature in the 3-$\micron$ band. 
Note that the signal of this object observed with AKARI was faint 
($\sim 9$~mJy in 3~$\micron$, or APmag = 13.35) and no significant spectral feature was observed above the detection limit. 

\item{419 Aurelia}\\
Aurelia is a C-type asteroid (or F-type in the Tholen taxonomy). 
The signal observed by AKARI was faint
($\sim 6$~mJy in 3~$\micron$, or APmag = 13.73). 
It has an absorption feature at around 2.8~$\micron$, but it was too noisy and below the detection limit. 

\footnotetext[5]{Nesvorny, D., 2012, NASA Planetary Data System, EAR-A-VARGBDET-5-NESVORNYFAM-V2.0}
\addtocounter{footnote}{1}

\item{423 Diotima}\\
Follow-up observations in 1--2.5~$\micron$ were performed by~\cite{Hasegawa2017}, and this object was classified 
as C-type according to the Bus-DeMeo taxonomy. 
This object is a member of 
the Eos collisional family (D. Nesvorny 2012\footnotemark[5], not in D. Nesvorny 2015\footnotemark[4]), 
but is likely to be an interloper based on a spin state analysis~\citep{Hanus2018}.
\cite{1990Icar...88..172J} conducted 3-$\micron$ band observations with IRTF and detected
no signal associated with hydrated minerals and/or water ice. 
Our AKARI results show an absorption feature at around 2.79~$\micron$, 
but its spectral shape is shallower and wider than that of typical hydrated minerals found in other 
asteroids in this study. 

\item{451 Patientia}\\
Patientia is a C-type asteroid, or Cb-type in the Bus taxonomy~\citep{Lazzaro2004}, it has a very flat light curve, indicating a spherical body~\citep{Michalowsk2005}.
\cite{2012Icar..219..641T} classified the 3-$\micron$ band spectral feature of this object into the Europa-like group. 
Our AKARI results show absorption features at around 3.06~$\micron$, which are classified as the 3-$\micron$ dent. 
A spectral feature appears at around 2.78~$\micron$, but is below the detection limit. 

\item{511 Davida}\\
This object is a C-type asteroid and a member of the Meliboea family (D. Nesvorny 2015\footnotemark[4]). 
A shape model of this object was given by~\cite{Conrad2007} based on Keck AO observations, and its
volume and bulk density are discussed by~\cite{Viikinkoski2017}. 
Its bulk density ($2.1 \pm 0.4$ g~cm$^{-3}$) suggests 
some degree of differentiation within the interior of the object. 
\cite{2012Icar..219..641T} classified the 3-$\micron$ band spectral feature of this object into the sharp group. 
Our AKARI results show absorption features at around 2.73~$\micron$ and, to a lesser degree, the 3.1-$\micron$ feature, which is below the detection limit. 
Its spectral shape is classified into the sharp group. 
\end{description}

\item B- and Cb-type asteroids
\begin{description}

\item{2 Pallas}\\
Pallas is a B-type asteroid and the third largest asteroid in the main-belt region; it is associated with a collisional family~\citep{Lemaitre1994} that resulted from a cratering event.
It is reported that the albedo and spectral properties of the Pallas collisional family members differ from 
those of other B-types~\citep{AliLagoa2013,AliLagoa2016}. 
\cite{Schmidt2012} discussed that water may play an important role in 
the thermal-physical evolution of this object. 
Surface heterogeneity of this object was detected with 
ground-based telescopes and Hubble observations \citep{Schmidt2009,Carry2010}. 
Our AKARI results show a significant feature found at around 2.74~$\micron$, which 
appears similar to the sharp group in \cite{2012Icar..219..641T}. 
\cite{Rivkin2015} discussed that the spectral shapes of Ch-type asteroids have the same spectral shape as that of 
2~Pallas; i.e., ``Pallas-type'' spectral group, which is consistent with the presence 
of phyllosilicates. 
In the AKARI observations, it is sufficiently bright and (partly) saturates the observed signal 
on the detector as described in section~\ref{sec:saturation}, while 
no unusual behavior is found in the spectra compared with the literature. 

\item{704 Interamnia}\\
Interamnia is a Cb-type asteroid with a diameter of 320~km. 
It is one of the largest bodies in the main-belt, but is not associated with a family~\citep{Rivkin2014}.
In the Tholen taxonomy, it is classified as F-type. 
Polarimetric observations suggest that this object has peculiar
surface properties~\citep{Belskaya2005}. 
\cite{2012Icar..219..641T} classified the 3-$\micron$ band spectral feature of this object into the sharp group. 
Our AKARI results also show absorption features at around 2.74~$\micron$, which is classified as the sharp group.

\end{description}

\item Cgh- and Ch-type asteroids
\begin{description}

\item{51 Nemausa}\\
Nemausa is a Cgh-type asteroid (or Ch-type in the Bus taxonomy). 
\cite{Fornasier1999} reported that this object displays an aqueous alteration signature according to ground-based observations 
at visible wavelengths. 
\cite{2015AJ....150..198R} reported a strong absorption feature for this object in the 3-$\micron$ band.  
Our AKARI results also show a significant absorption feature at around 2.77~$\micron$, which 
belongs to the sharp group. 
The band depth at 2.7~$\micron$ for this object is the strongest among all targets in this study ($\sim$ 47\%). 

\item{106 Dione}\\
Dione is a Cgh-type asteroid, for which \cite{2003MPS...38.1383R} reported evidence of hydration in the 3-$\micron$ band with a ground-based
telescope. 
Our AKARI results show a significant absorption feature at around 2.76~$\micron$, which 
is classified as the sharp group.

\item{13 Egeria}\\
Egeria is a Ch-type asteroid.
\cite{Burbine1998} discussed that the band shape at around 0.7~$\micron$ is associated with CM chondrites. 
\cite{2003MPS...38.1383R} reported evidence of hydration in the 3-$\micron$ band with a ground-based
telescope. 
\cite{2015Icar..257..185T} classified the 3-$\micron$ band spectral feature of this object into the sharp group, which is 
confirmed by our AKARI results to be at around 2.76~$\micron$. 

\item{49 Pales}\\
Pales is a Ch-type asteroid and a slow rotator, taking 20.70~hr for one revolution~\citep{Warner2009}. 
No studies to date report observations in the 3-$\micron$ band; however, our AKARI results show an absorption feature at around 2.75~$\micron$, which 
is classified into the sharp group.

\item{50 Virginia}\\
Virginia is a Ch-type asteroid or X-type in the Tholen taxonomy, Virginia has no observations in the 3-$\micron$ band so far. 
Conversely, our AKARI results show an absorption feature at around 2.74~$\micron$, which 
is classified into the sharp group; however, the signal observed with AKARI is faint 
($\sim 9$~mJy in 3~$\micron$, or APmag = 13.21). 

\item{121 Hermione}\\
Hermione is a Ch-type asteroid in the Cybele group, Hermione is a binary system with a moon S/2002 (121) 1 with a diameter of 
$\sim$ 12 km~\citep{Marchis2006}, which converts to a diameter ratio for this system of 0.06. 
\cite{2012Icar..219..641T} classified the 3-$\micron$ band spectral feature of this object into the sharp group. 
\cite{2012Icar..221..453H} found a deep 3-$\micron$ absorption feature with IRTF observations and 
no 10-$\micron$ emission feature with Spitzer/IRS. 
Our AKARI results also show an absorption feature at around 2.78~$\micron$, which is classified into the sharp group. 
Two observations were performed with AKARI at 9.9-h intervals; no significant difference is observed. 

\item{127 Johanna}\\
\cite{Hasegawa2017} reported that this object is classified as Ch-type in the Bus-DeMeo taxonomy and \cite{2015AJ....150..198R} reported an absorption feature of this object in the 3-$\micron$ band. 
The signal for this object observed with AKARI is faint 
($\sim 5$~mJy in 3~$\micron$, or APmag = 13.91). 
One feature appears at around 2.81~$\micron$, which is classified into the sharp group, but
it is almost indiscernible from the noise. 

\item{145 Adeona}\\
Adeona is a Ch-type asteroid and the largest member of the asteroid family bearing its name (D. Nesvorny 2015\footnotemark[4]).
\cite{Fornasier1999} reported an aqueous alteration signature according to ground-based observations 
at visible wavelengths. 
The signal according to AKARI is faint
($\sim 9$~mJy in 3~$\micron$, or APmag = 13.03). 
One feature appears at around 2.76~$\micron$, which is classified into the sharp group. 

\end{description}

\end{itemize}

\subsubsection{X-complex asteroids}

We divide the X-complex into three groups according to their albedo (c.f., \citealt{Tholen1984}): 
low-albedo ($p_\mathrm{v} < 0.1$), 
moderate-albedo ($0.1 < p_\mathrm{v} < 0.3$), 
and high-albedo ($p_\mathrm{v} > 0.3$). 
Some low- and moderate-albedo X-complex asteroids have an absorption feature in the 2.7-$\micron$ band, 
similar to the C-complex asteroids. 

\begin{itemize}
\item{Low-albedo X-complex asteroids}
\begin{description}

\item{46 Hestia}\\
Hestia is an Xc-type asteroid (or P-type in the Tholen taxonomy), Hestia is a slow rotator, with one revolution lasting 21.04~hr~\citep{Warner2009}. 
There are no reports to date of features in the 3-$\micron$ band. 
\cite{Fieber-Beyer2015} discussed the relationship between these objects and CR chondrites based on 
the vis-NIR observations. 
Our AKARI results show absorption features at around 2.74~$\micron$, which is classified as the sharp group, and some features at around 3.1~$\micron$, 
but the spectral shape appears bumpy. 
The observed signal is faint ($\sim 12$~mJy in 3~$\micron$, or APmag = 13.42) and its spectrum 
suffered from noise.
Two observations were performed with AKARI at 100-min intervals with no significant difference found between them.

\item{56 Melete}\\
Melete is an Xk-type asteroid (or P-type in the Tholen taxonomy). 
No features have yet been observed in the 3-$\micron$ band. 
Our AKARI results show absorption features at around 2.73~$\micron$ (sharp).  
There is also a small feature at around 3.1~$\micron$, which is below the detection limit. 

\item{65 Cybele}\\
Cybele, an Xk-type asteroid (or P-type in the Tholen taxonomy) 
with a diameter of 300 km, belongs to the Cybele group bearing its own name, which is located 
in the outer main-belt region. 
\cite{2011AA...525A..34L} made spectroscopic observations with IRTF and found 
an absorption band centered at 3.10~$\micron$, which is associated with water ice as frost, but no feature associated with hydrated silicates in the 2-4~$\micron$ band. The former feature is similar to that of C-type asteroid 24~Themis. 
\cite{2012Icar..219..641T} classified the 3-$\micron$ band spectral feature of this object into the rounded group. 
Our AKARI results also show an absorption feature at around 3.1~$\micron$, but its spectral shape is 
sharper than previous works. 

\item{87 Sylvia}\\
Sylvia is an X-type asteroid (or P-type in the Tholen taxonomy), 
Sylvia is a triple system with two moons, Remus and Romulus, 
with respective diameters of $\sim$ 7~km and $\sim$ 18~km~\citep{Marchis2005}. 
The respective diameter ratios of this system are 0.03 and 0.07. 
There are no previous reports of spectroscopic observations in the 3-$\micron$ band. 
The signal of this object observed with AKARI is faint 
($\sim 13$~mJy in 3~$\micron$, or APmag = 13.30); it shows some features in the 3-$\micron$ band, but they are below the detection limit. 

\item{140 Siwa}\\
Siwa is an Xc-type asteroid (or P-type in the Tholen taxonomy) 
and a slow rotator, taking 34.45~hr for one revolution~\citep{Warner2009}.  
No absorption features were found in previous studies~\citep{2012Icar..219..641T}; i.e., it is a featureless object.
Our AKARI results also show no significant features in the 3-$\micron$ band. 

\item{153 Hilda}\\
Hilda is an X-type asteroid (or P-type in the Tholen taxonomy) 
and the largest member of the asteroid family bearing its name (D. Nesvorny 2015\footnotemark[4]).
\cite{2012Icar..219..641T} classified the 3-$\micron$ band spectral feature of this object into the rounded group. 
The signal observed by AKARI is faint 
($\sim 7$~mJy in 3~$\micron$, or APmag = 14.01); thus, we cannot identify any significant features with a sufficient S/N ratio. 

\item{173 Ino}\\
Ino is an Xk-type asteroid 
and the largest member of the asteroid family bearing its name (D. Nesvorny 2015\footnotemark[4]). 
\cite{1990Icar...88..172J} reported some hydration of this object. 
Our AKARI results show an absorption feature at around 2.75~$\micron$, which 
is classified into the sharp group. 
A small dent shape of uncertain origin appears at around 2.97~$\micron$.

\item{336 Lacadiera}\\
Lacadiera is an Xk-type asteroid (or D-type in the Tholen taxonomy). 
There are no previous reports of hydration detected in the 3-$\micron$ band (e.g., \citealt{2002Icar..156..434C}).
The signal observed by AKARI is too faint 
($\sim 4$~mJy in 3~$\micron$, or APmag = 14.24); thus, no significant features were observed with a sufficient S/N ratio. 

\item{476 Hedwig}\\
Hedwig is an Xk-type asteroid (or P-type in the Tholen taxonomy), Hedwig is a slow rotator, with one revolution lasting 27.33~hr~\citep{Warner2009}. 
No spectroscopic observations have been made in the 3-$\micron$ band. 
The AKARI signal is faint 
($\sim 11$~mJy in 3~$\micron$, or APmag = 13.42); however, this object has a sharp spectral feature at around 2.75~$\micron$, which 
is typically found in C-complex asteroids in AKARI data. 

\end{description}

\item{Moderate-albedo X-complex asteroids}
\begin{description}

\item{16 Psyche}\\
Psyche is an Xk-type asteroid 
(or M-type in the Tholen taxonomy). 
The {\it Psyche} mission was selected by the NASA Discovery program to 
rendezvous with asteroid 16~Psyche~\citep{Elkins-Tanton2017}. 
Radar albedo~\citep{Ostro1985,Magri2007,Shepard2008,Shepard2017}, 
thermal inertia~\citep{Matter2013}, and density estimation~\citep{Kuzmanoski2002} 
indicate that this asteroid has a predominantly metal-rich surface. 
There is no dynamical family associated with this object~\citep{Davis1999}. 
\cite{2017AJ....153...31T} detected a 3-$\micron$ band feature at approximately a $\sim 3\%$ level, 
possibly associated with water or hydroxyl, which is proposed to have exogenic origins 
(e.g., \citealt{Landsman2017, Avdellidou2018}). 
In contrast, our AKARI data indicate no significant absorption feature in the 3-$\micron$ band. 
Note that a $\sim$ 4\% band depth at around 3.1~$\micron$ is found in our data, but it is below the detection limit. 
Two observations were performed with AKARI at 100-min intervals with no difference 
found between them. 

\item{21 Lutetia}\\
Lutetia is an Xc-type asteroid
(or M-type in the Tholen taxonomy), its bulk density was determined from the spacecraft Rosetta flyby as 
$3.4 \pm 0.3$~g~cm$^{-3}$~\citep{Patzold2011}. 
Rosetta also performed spectroscopy of Lutetia~\citep{2011Sci...334..492C} and reported 
no absorption features in the spectral range from 0.4 to 3.5~$\micron$. 
IRTF observations~\citep{2011Icar..216..650V} also revealed absence of 3-$\micron$ features. 
On the other hand, \cite{2011Icar..216...62R} reported the detection of a 3-$\micron$ band feature 
at 3--5\% level in the southern hemisphere, the other side of the asteroid which is not visible 
to Rosetta \citep{Rivkin2015AIV, Barucci2017}. 
Our AKARI results show no absorption features in the 3-$\micron$ band. 
A small dent shape with a $\sim$ 2\% band depth appears at around 2.7~$\micron$, 
which might come from contamination due to background stars. 

\item{22 Kalliope}\\
Kalliope is an X-type asteroid 
(or M-type in the Tholen taxonomy) 
with an albedo of 0.24~\citep{Usui2011} and bulk density of 3.35~g~cm$^{-3}$~\citep{Descamps2008}, 
Kalliope is a binary system with a moon, Linus (22~Kalliope I), with a diameter of 28~km and an orbital period of 3.5954 days~\citep{Descamps2008}.
The diameter ratio of this system is 0.2. 
\cite{2000Icar..145..351R} showed evidence for hydrated minerals on Kalliope with a $\sim$ 10\% band depth 
in the 3-$\micron$ band based on IRTF observations. 
Our AKARI results show a small feature with a $\sim$ 4\% band depth at around 2.78~$\micron$. 
Two observations were performed with AKARI at 3.3-hr intervals with no significant difference. 

\item{55 Pandora}\\
Pandora is an Xk-type asteroid 
(or M-type in the Tholen taxonomy)
with a relatively high albedo: 0.34 observed by AKARI~\citep{Usui2011} or 0.20 according to WISE~\citep{Masiero2012}. 
\cite{1990Icar...88..172J} reported an absorption feature in the 3-$\micron$ band, and 
\cite{2000Icar..145..351R} also 
showed a hydration feature with a band depth of $\sim$ 9\% and temporal variation in this spectral feature. 
Our AKARI results show a feature at around 2.8~$\micron$, which is below the detection limit. 
Three observations were performed with AKARI within 6.6~hr but all of them  
suffered from contamination due to background stars. 

\item{69 Hesperia}\\
Hesperia is an Xk-type asteroid
(or M-type in the Tholen taxonomy).  
\cite{2015Icar..252..186L} reported a 3-$\micron$ absorption feature with a band depth of 6.2\%. 
Our AKARI results show a feature at around 2.8~$\micron$ that is below the detection limit. 
Note that only one pointed observation was conducted for this object with AKARI. 

\item{92 Undina}\\
Undina is an Xk-type asteroid. 
\cite{2000Icar..145..351R} 
 reported a 3-$\micron$ absorption feature with a band depth of 9.3\%. 
Our AKARI results show a small feature with a $\sim$ 6\% band depth at around 2.76~$\micron$. 

\item{129 Antigone}\\
Antigone is an Xk-type asteroid
(or M-type in the Tholen taxonomy).  
\cite{2000Icar..145..351R} 
 reported a 3-$\micron$ absorption feature with a band depth of 14.4\%. 
Our AKARI results show a peculiar wavy structure 
related to contamination by background stars 
(Kmag=13.5 in the data of ObsID:1520147-001, and Kmag=15.7 in the data of ObsID:1520148-001).

\item{135 Hertha}\\
Hertha is an Xk-type asteroid
(or M-type in the Tholen taxonomy),  
Hertha is one of the largest members in 
the Nysa-Polana complex (e.g., \citealt{Dykhuis2015}). 
\cite{2000Icar..145..351R} 
reported a 3-$\micron$ absorption feature with a band depth of 10.2\% 
and 
\cite{Rivkin2002} discussed the 0.7~$\micron$ absorption of this object. 
Our AKARI results show a peculiar wavy structure of uncertain origin, which is below the detection limit. 

\item{161 Athor}\\
Athor is an Xc-type asteroid
(or M-type in the Tholen taxonomy).  
\cite{2000Icar..145..351R} reported absence of 3-$\micron$ features. 
The signal observed with AKARI is too faint 
($\sim 12$~mJy in 3~$\micron$, or APmag = 13.31)
to identify any significant feature with a sufficient S/N ratio. 

\item{216 Kleopatra}\\
Kleopatra is an Xe-type asteroid (or M-type in the Tholen taxonomy), 
which is famous for its peculiar bilobate, or ``dog-bone'' shape~\citep{Ostro2000}. 
This is a triple system with two moons, 
Cleoselene and Alexhelios, with diameters of $\sim$ 7~km and 
$\sim$ 9~km, respectively~\citep{Descamps2011}. Diameter ratios of this system are 0.06 and 0.07. 
\cite{2015Icar..252..186L} showed 
evidence for hydrated minerals in Kleopatra using IRTF observations; approximately 5\% on average, as well as rotational variability in the depth of its 3-$\micron$ feature. 
Our AKARI results show a small feature with a $\sim$ 3\% band depth at around 2.78~$\micron$. 
Three observations were performed with AKARI; 
the first two were separated by 5-hr intervals and the last was conducted 1.46~years after the first two 
because of observational scheduling. Among these three, no significant spectral differences are observed above the noise level. 

\item{250 Bettina}\\
Bettina is an Xk-type asteroid
(or M-type in the Tholen taxonomy). 
\cite{Vernazza2009} discussed a similarity of spectral shape between this object and 
mesosiderite, a stony-iron meteorite, at 0.4--2.5~$\micron$. 
No observations exist in the 3-$\micron$ band. 
The signal observed by AKARI is faint 
($\sim 13$~mJy in 3~$\micron$, or APmag = 13.31) and no significant absorption feature is found in the 3-$\micron$ band. 
One data (ObsID:1520161-001) is removed by contamination so 
only one observational data is used for this object. 

\end{description}

\item{High-albedo X-complex asteroids}
\begin{description}

\item{44 Nysa}\\
Nysa is an Xn-type asteroid 
(or E-type in the Tholen taxonomy) 
with a high albedo of 0.48 \citep{Usui2011, Masiero2012},  
Owing to its high albedo, the contribution of thermal emission is small at 2.5--5~$\micron$. 
It is the largest member of the Nysa-Polana family (D. Nesvorny 2015\footnotemark[4]). 
\cite{1995Icar..117...90R} detected the 3-$\micron$ band depth on Nysa with 14\%, 
which is attributed to hydrated minerals. 
By removing one data (ObsID:1520189-001) due to contamination as described above, 
only one observational data from AKARI is used for this object. 
It shows a  peculiar wavy structure of uncertain origin. 
An unidentified feature is also observed with a $\sim$ 14\% band depth at around 3.08~$\micron$. 

\item{64 Angelina}\\
Angelina is an Xe-type asteroid (or E-type in the Tholen taxonomy) 
with a high albedo; 0.52 according to AKARI~\citep{Usui2011} or 0.48 for WISE~\citep{Masiero2014}. 
Owing to its high albedo, the thermal emission contribution is small at 2.5--5~$\micron$. 
There are no reports of observations in the 3-$\micron$ band. 
The signal observed by AKARI is faint 
($\sim$ 12-13~mJy in 3~$\micron$, or APmag = 12.83) and its spectrum is flat with no features 
in the 3-$\micron$ band. 

\end{description}

\end{itemize}

\subsubsection{D-complex asteroids}

D-complex asteroids are darker and located further from the observer; thus, these asteroids are fainter (APmag $> 13$). 
One has an absorption feature in the 2.7-$\micron$ band, like C-complex asteroids, and the other two do not. 

\begin{description}

\item{773 Irmintraud}\\
Irmintraud is a T-type asteroid, and there are a couple of reports suggesting the existence of water on this object 
(e.g., \citealt{1990Icar...83...16L}, \citealt{1990Icar...88..172J}, \citealt{Howell1995}, \citealt{Merenyi1997}).
\cite{2003GeoRL..30.1909K} performed near-infrared photometric and spectroscopic observations of this object
and concluded that the Tagish Lake meteorite is related to D-type asteroids. 
The signal observed by AKARI is faint 
($\sim 5$~mJy in 3~$\micron$, or APmag = 14.36) and the data quality is not sufficient to interpret any pattern in the spectral shape at 2.5--3.5~$\micron$, which is almost hidden by noise; thus, 
no significant features are identified with a sufficient S/N ratio. 

\item{308 Polyxo}\\
Polyxo is a T-type asteroid. 
\cite{Rivkin2002} reported that most T-types have a 3-$\micron$ band feature. 
\cite{HiroiHasegawa2003} made spectral fitting between Polyxo with ground-based observations and the Tagish Lake meteorite. 
\cite{2012Icar..219..641T} classified the 3-$\micron$ band spectral feature of this object into the sharp group. 
The AKARI signal for this object is faint ($\sim 13$~mJy in 3~$\micron$, or APmag = 13.34); however, it has a sharp spectral feature with a $\sim$ 15\% band depth 
at around 2.76~$\micron$, which is typically found in C-complex asteroids in AKARI data. 

\item{361 Bononia}\\
Bononia is a D-type asteroid and belongs to the Hilda group. 
\cite{2012Icar..219..641T} classified the 3-$\micron$ band spectral feature of this object into the rounded group. 
The AKARI signal of this object is faint 
($\sim 6$~mJy in 3~$\micron$, or APmag = 14.26). 
A pattern is observed in the spectral shape at 2.5--3.5~$\micron$, which is almost indistinguishable from noise; thus, no significant features have a sufficient S/N ratio. 

\end{description}

\subsubsection{S-complex asteroids}
Our AKARI results show that 
only a few S-complex asteroids have an absorption feature with a few percent band depth 
in the 3-$\micron$ band. 

\begin{description}

\item{5 Astraea}\\
Astraea is an S-type asteroid.
\cite{1990Icar...88..172J} observed Astraea with IRTF and 
reported no absorption feature in the 3-$\micron$ band. 
Our AKARI results also show no significant feature in the spectrum of this object. 

\item{6 Hebe}\\
Hebe is an S-type asteroid.
\cite{2001LPI....32.1723R} detected a 
3-5\% band depth in the 3-$\micron$ band of this object with UKIRT. 
This level of feature is below the detection limit of AKARI observations, and our results indicate no significant feature in this object. 

\item{7 Iris}\\
Iris is an S-type asteroid.
\cite{1997PhDT........10R} did not detect any hydrated minerals on this object with IRTF. 
Our AKARI results also show no significant feature in the spectrum of this object in the 3-$\micron$ band.

\item{33 Polyhymnia}\\
Polyhymnia is an S-type asteroid. 
No previous observations have been reported for the 3-$\micron$ band. 
Our AKARI results show a small feature with a $\sim$ 3\% band depth at around 2.68~$\micron$. 

\item{40 Harmonia}\\
Harmonia is an S-type asteroid, and our AKARI results show no significant feature in the spectrum of this object in the 3-$\micron$ band.

\item{79 Eurynome}\\
Eurynome is an S-type asteroid.
Our AKARI results show a slightly wavy structure with noise, but no significant feature in the 3-$\micron$ band.

\item{148 Gallia}\\
Gallia is an S-type asteroid and 
the largest member of the asteroid family bearing its name (D. Nesvorny 2015\footnotemark[4]).
Gallia is a slow rotator, taking 20.66~hr for one revolution~\citep{Warner2009}. 
\cite{1990Icar...88..172J} observed 
Gallia with IRTF and 
reported no absorption feature in the 3-$\micron$ band. 
Our AKARI results show a wavy structure with noise, but no significant feature in the 3-$\micron$ band. 

\item{371 Bohemia}\\
Bohemia is an S-type asteroid. 
The signal observed by AKARI is faint ($\sim 6$~mJy in 3~$\micron$, or APmag = 13.89) 
and the data quality is insufficient to indicate a pattern in the spectral shape at 2.5--3.5~$\micron$, which is almost hidden by noise; thus, no significant features are identified with a sufficient S/N ratio. 

\item{532 Herculina}\\
Herculina is an S-type asteroid. 
\cite{1990Icar...88..172J} observed 
Herculina with IRTF and reported no absorption feature in the 3-$\micron$ band. 
Our AKARI results also show no significant feature in the spectrum of this object in the 3-$\micron$ band.

\item{8 Flora}\\
Flora is an Sw-type asteroid 
and has a high spectral slope at 0.45--2.45~$\micron$. 
\cite{1983Icar...55..245E} observed 
Flora with UKIRT and found a steep slope of the spectrum at 3--4~$\micron$. 
This spectral shape was considered a continuation of that in shorter wavelengths. 
Our AKARI data also has a positive spectral slope of $\mathcal{S}=0.102~\micron^{-1}$, which is higher than 
the mean value of S-complex asteroids. 
A small dent shape of uncertain origin with a $\sim$ 6\% band depth appears at around 3.09~$\micron$. 
Note that only one observation was conducted for this object with AKARI. 

\item{89 Julia}\\
89~Julia is an Sw-type asteroid, the largest member of the asteroid family bearing its name (D. Nesvorny 2015\footnotemark[4]), 
and considered a parent body of the quasicrystal-bearing CV meteorite Khatyrka~\citep{Meier2018}. 
Our AKARI results show a slightly wavy structure, but no significant feature in the 3-$\micron$ band.
It has a large positive spectral slope from 2.5~$\micron$ to 3.5~$\micron$ ($\mathcal{S} = 0.112~\micron^{-1}$). 

\item{246 Asporina}\\
Asporina is an A-type asteroid, this object is considered a member of the rare olivine-dominated asteroids (e.g., \citealt{Sanchez2014}). 
In our AKARI data, 
a small dent shape with a $\sim$ 9\% band depth appears at around 3.08~$\micron$; however, the AKARI signal is faint 
($\sim 11$~mJy in 3~$\micron$, or APmag = 13.82) 
and the data quality is not sufficient for further analysis.

\item{354 Eleonora}\\
Eleonora is an A-type asteroid.
There is a possibility that the surface materials of this object contain a pallasite assemblage~\citep{Gaffey2015}.
In our AKARI data, 
a small dent shape of uncertain origin with a $\sim$ 4\% band depth appears at around 3.06~$\micron$. 

\item{349 Dembowska}\\
Dembowska is an R-type asteroid with a near-infrared spectrum of two strong absorption 
features at 1 and 2~$\micron$~\citep{Bus2002, DeMeo2009}, Dembowska has a high albedo of 0.31~\citep{Yu2017}. 
In our AKARI data, 
small dent shapes appear at around 2.79~$\micron$ (4\%) and 3.09~$\micron$ (6\%), 
which appear in two pointed observations. 
A highly negative slope ($\mathcal{S}=-0.033~\micron^{-1}$) of this spectrum is observed at 2.5--4~$\micron$, 
which might be characteristic of R-type, olivine-rich basaltic asteroids \citep{Leith2017}. 

\item{9 Metis}\\
Metis is an L-type asteroid (or S-type in the Tholen taxonomy).
Our AKARI results show a bumpy structure at around 2.9~$\micron$, which might be attributable to be contamination by background stars. 
These are below the detection limit; thus, no obvious features are found in our AKARI data. 

\item{387 Aquitania}\\
Aquitania is an L-type asteroid (or S-type in the Tholen taxonomy),
Aquitania is a slow rotator, taking 24.14~hr for one revolution~\citep{Warner2009}. 
From polarimetric observations~\citep{Masiero2009},
it is considered a member of the so-called Barbarians~\citep{Cellino2006}, 
which represent some of the oldest surfaces in the solar system. 
Our AKARI results show 
a bumpy structure at around 3.1~$\micron$, which might come from contamination due to background stars. 
These are below the detection limit, and thus no obvious features are found in our AKARI data. 

\item{42 Isis}\\
Isis is a K-type asteroid (or S-type in the Tholen taxonomy). 
An unidentified feature with a $\sim$ 8\% band depth is observed at around 3.09~$\micron$
in AKARI data. 

\end{description}

\section{Discussion}

\subsection{Advantages and limitations of the AKARI spectroscopic observations}
\begin{figure}
\begin{center}
\includegraphics[width=140mm]{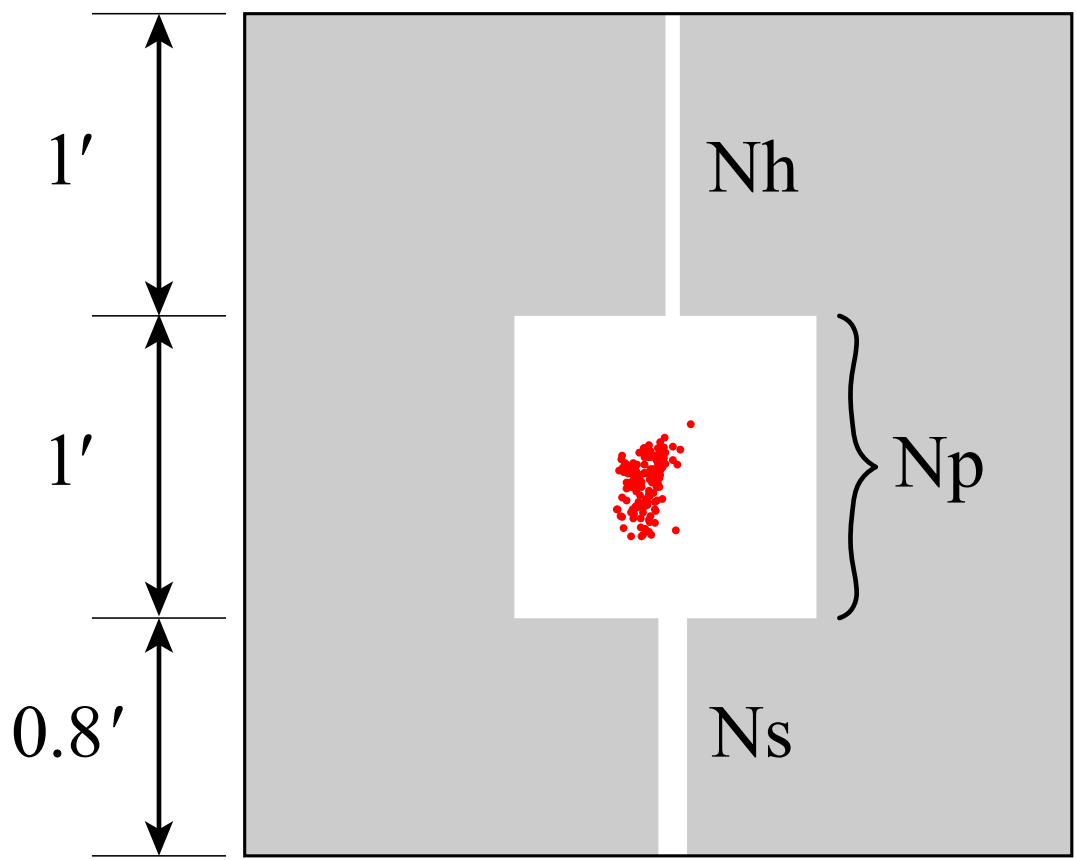}
\end{center}
\caption{
Distribution of the target positions in the ``Np'' window in the reference image 
(see also figure~\ref{fig:data-example}). 
The gray area is a masked region on the detector. 
}\label{fig:Np window target position}
\end{figure}

The greatest advantage of the AKARI observations is 
free from disturbance by telluric absorption and thus able to obtain spectra continuously 
from 2.5~$\micron$ to 5~$\micron$, which can fully cover the peak wavelength of the 2.7-$\micron$ band.  
On the other hand, the effective aperture size of the AKARI telescope is 68.5~cm and the exposure time of 
each pointed observation is only 10~min in total, which is limited by severe constraints of the attitude control. 
Thus the present survey provides a flux-limited sample. 
The detection sensitivity is $\sim$ 1.3 mJy at 3~$\micron$ with 10$\sigma$ (see the IRC Data User Manual). 
Note that the sensitivity during Phase 3, the warm mission phase, 
is likely worse than this because of the temperature change
due to degradation of the cryocooler~\citep{Onaka2010}. 
Specifically, the number of hot pixels on the detector increases in Phase 3. 
To constitute data redundancy, data from two or three pointed observations are averaged to generate reflectance spectra 
in this study. 
Spectroscopic observations in the ``Np'' window are a type of slitless spectroscopy.
The distribution of the target positions in the ``Np'' window in the reference image is shown in 
figure~\ref{fig:Np window target position}. 
This ``Np'' window is effective for observations of point sources, but can be vulnerable to contamination by 
neighboring stars and ghost images due to nearby bright sources.  
Our observations were scheduled to avoid the region at around the galactic plane, i.e., our targets were 
located at galactic latitudes of $|b| > 16^{\circ}$.
Insufficient flat-fielding is also an issue for the present dataset (see section~\ref{sec:Data Reduction}). 

\subsection{Identification of the 3-$\micron$ band shape}
In this study, the 3-$\micron$ band shape is classified into sharp, w-shape, and 3-$\micron$ dent (table~\ref{tab:band_depth}). This is determined by a combination of $\mathcal{D}_{2.7}$ and 
$\mathcal{D}_{3.0}$ described in section~\ref{sec:General trend of the spectra}. 
This criterion assumes that asteroid spectra contain no narrow or 
steeply peaking features like atomic or molecular line spectra, but 
have broad features of typically $\sim$ 0.1~$\micron$ or wider in the 2.7- or 3.1-$\micron$ band. 

Classification of the 3-$\micron$ band shape is not yet fully established 
compared to the taxonomies based on 
the vis-NIR spectra 
(e.g., \citealt{Tholen1984, Bus2002, Lazzaro2004, DeMeo2009}). 
\cite{2012Icar..219..641T} and \cite{2015Icar..257..185T} classified 
3-$\micron$ band shapes into sharp, rounded, Europa-like, and Ceres-like (hereafter Takir class), 
while \cite{Rivkin2012} classified them into Pallas-like, Ceres-like, Themis-like, and Lutetia-like (hereafter Rivkin class). 
These classifications are summarized in table~\ref{tab:3um group}. 
Our classification is generally consistent with Takir class and Rivkin class. 
Observations with the space telescope in this study detected the 2.7-$\micron$ band feature 
without atmospheric disturbance, which is a big help to classify asteroids into groups of 
the 3-$\micron$ band spectral shape. 

There are five asteroids among our targets that are classified into the sharp group of the Takir class: 
13~Egeria (Ch), 121~Hermione (Ch), 308~Polyxo (T), 511~Davida (C), and 704~Interamnia (Cb). 
\cite{2012Icar..219..641T} reported that this spectral shape is most consistent with the presence 
of hydrated silicates. Certainly, our AKARI results for these objects show significant absorption 
features at 2.7~$\micron$. It should be noted that small features appear 
at around 3.1~$\micron$, which might be obscured by the strong features at 2.7~$\micron$. 
There are also four asteroids among our targets that are classified into the rounded-shape of the Takir class:
24~Themis (C), 65~Cybele (Xk), 153~Hilda (X), and 361~Bononia (D). 
Based on AKARI data, 24~Themis has a wider absorption feature at both 2.7~$\micron$ 
and 3.1~$\micron$ and is thus classified into the w-shape in this study. 
153~Hilda and 361~Bononia were not observed with a sufficient S/N with AKARI. 
Ceres-like objects include 1~Ceres (C) itself and 10~Hygiea (C). 
These objects display two clear features at 2.7~$\micron$ and 3.1~$\micron$ with comparable band depths, 
which belong to the w-shape. 
Europa-like objects include 52~Europa (C) itself and 451~Patientia (C). 
Europa seems to have two features at 2.7~$\micron$ and 3.1~$\micron$, but the former is too weak to be classified 
as the w-shape. Patientia also has a low S/N spectrum in our data. 

In general, 
it is difficult to accurately classify an object into the featureless category. 
To make reliable classification, it is necessary to ensure that no features exist 
within a certain level of accuracy, which requires high S/N data throughout the wavelength of interest. 
The reflectance spectra obtained in this study include remnant wavy structures on a level of a few percent; thus, 
it is difficult to distinguish real features from these structures. 
There are three candidates for featureless spectra in the AKARI data, which have a band depth of less than 2\% 
at 3-$\micron$: 6~Hebe (S-type) 140~Siwa (Xc-type), and 532~Herculina (S-type). 
Note that 140~Siwa is classified as featureless according to the Takir class
\citep{2012Icar..219..641T}, which 
does not show any feature above their noise level in the 3-$\micron$ band. 
21~Lutetia is reported to have a broad absorption feature in the 3-$\micron$ band in the Rivkin class 
unlike Pallas, Ceres, or Themis \citep{2011Icar..216...62R}. 
Our AKARI data, on the other hand, 
is truncated at $\lambda_\mathrm{trunc} = 3.57~\micron$, and thus the broad spectral shape cannot be fully covered. 
Featureless spectra can constrain the abundance of surface materials, 
especially once they are combined with other spectral data. For example, 
\cite{Emery2003} reported featureless spectra of 20 Trojan asteroids 
in the 3-$\micron$ region observed with IRTF and these featureless 
are attributed to the presence of anhydrous silicates \citep{Emery2004}.

\subsection{Peak wavelength in the 2.7-$\micron$ band feature}
\begin{figure}
\begin{center}
\includegraphics[width=140mm]{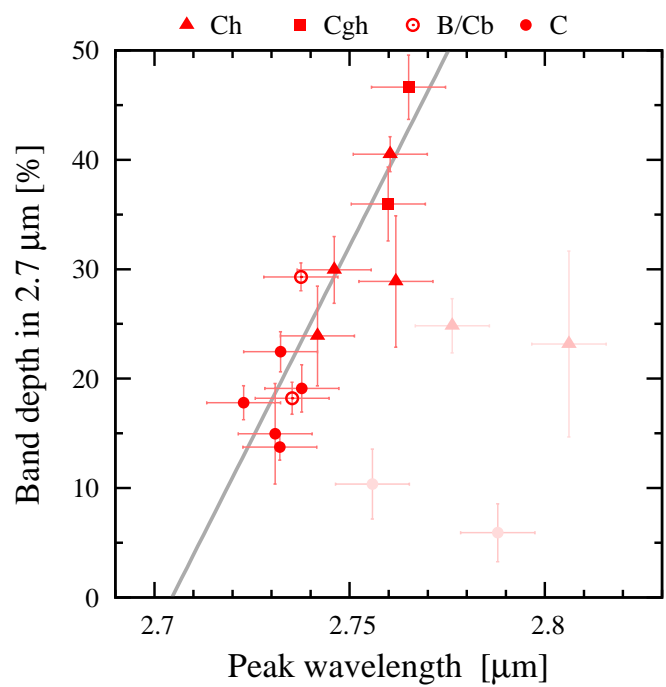}
\end{center}
\caption{
Relationship between the band depths at 2.7-$\micron$ against the peak wavelength for C-complex asteroids. 
Marks are the same as in figure~\ref{fig:orbital_elements}. 
The correlation coefficient is 0.88. 
The solid gray line indicates the fitted linear line. 
The thin red dots denote the asteroid treated as outliers for fitting (see text). 
}\label{fig:peak_fitting}
\end{figure}

As seen in figure~\ref{fig:strength-peakwavelength}, 
the peak wavelength of the 2.7-$\micron$ band feature is concentrated at around 2.75~$\micron$. 
In particular, 
C-complex asteroids have a trend between the peak wavelength and 
the band depth with S/N $> 2$. 
Figure~\ref{fig:peak_fitting} shows this 
trend for 17 C-complex asteroids. 
There are four outliers with longer peak wavelengths: 
24~Themis at 2.76~$\micron$, 
121~Hermione at 2.81~$\micron$, 127~Johanna at 2.85~$\micron$,
and 423~Diotima at 2.79~$\micron$. 
Except for these four, there is a correlation between the peak wavelength and the band depth
among 13 C-complex asteroids: 
\begin{eqnarray}
\mathcal{D}_{2.7} &=& %
705.7~(\lambda_\mathrm{peak} - 2.704)\ , 
\label{eq:depth-wavelength}
\end{eqnarray}
where $\mathcal{D}_{2.7}$ is in units of percent (\%) and 
$\lambda_\mathrm{peak}$ is the peak wavelength in units of~$\micron$.
The correlation coefficient is 0.88. 


The correlation should be used with caution because only 13 asteroids comprise this trend; thus, 
it may be affected by observational bias. 
Nevertheless, this correlation may be important for considering asteroid spectra in meteorite research. 
Based on heating experiments of meteorites in the laboratory, Yamashita et al. (in prep) reported a peak-wavelength shift 
in the 2.7-$\micron$ band of hydrated minerals because of the dehydration process. 
The band depth of the 2.7-$\micron$ band indicates an abundance of phyllosilicate, and 
the peak wavelength of the 2.7-$\micron$ band indicates the Mg/Fe ratio in phyllosilicate. 
During the dehydration process by heating, phyllosilicate decreases and the Mg/Fe ratio simultaneously increase because 
of the progressive replacement of the phyllosilicates interlayer cations Fe$^{2+}$ by Mg$^{2+}$ 
(e.g., \citealt{Rubin2007, Beck2010, Nakamura2017}). 
This leads to a decrease of the band depth along with a peak wavelength shift toward shorter 
wavelengths. Therefore, equation~(\ref{eq:depth-wavelength}) can 
be interpreted in terms of the meteorite dehydration history. 

Some reports detected hydrated minerals on the lunar surface using 
in-situ observations from spacecrafts: 
Cassini~\citep{Clark2009a}, Deep Impact~\citep{Sunshine2009}, and Chandrayaan-1~\citep{Pieters2009}. 
These observations found peak wavelengths near 2.8-3.0~$\micron$. 
Similar absorption was also detected in the lunar soil of Apollo~16 and 17~\citep{Ichimura2012}. 
It is considered that the mechanisms forming these lunar hydrated minerals are different from those on asteroids: 
they may be associated with interaction of the lunar regolith with solar wind proton implantation. 
These minerals appear to have absorption features near 2.8--3.0~$\micron$. 
On the other hand, phyllosilicate absorption caused by aqueous alteration in 
meteorites appears in the vicinity of 2.7~$\micron$ (e.g., \citealt{Rivkin2015AIV}). 
Features found in most asteroids observed with AKARI are likely associated with 
this aqueous alteration. 

\subsection{Spectral slope in the 3-$\micron$ band}
The spectral slope measured in the available wavelength range is shown in figure~\ref{fig:slope-strength}.
The mean value of the spectral slope of total 64 objects is 
$0.047 \pm 0.047~\micron^{-1}$, 
which is almost flat at these wavelengths, indicating that the thermal component subtraction described 
in section~\ref{sec:Thermal Component Subtraction} 
works well; 
if the subtraction was insufficient, it may have caused a significant slope toward longer wavelengths. 
The mean value of the slope of each complex is: 
$\overline{\mathcal{S}(\mathrm{C})}=0.000~\pm~0.031~\micron^{-1}$, 
$\overline{\mathcal{S}(\mathrm{X})}=0.063~\pm~0.041~\micron^{-1}$, 
and 
$\overline{\mathcal{S}(\mathrm{S})}=0.059~\pm~0.045~\micron^{-1}$. 
Spectral slope makes a major contribution to derive a taxonomic classification of asteroids
base on the vis-NIR spectral data
(e.g., \citealt{DeMeo2009}). 
Our results in this study do not conflict with the trend at shorter wavelengths, that is, 
C-complex asteroids have a general trend of the flat slope, and 
X-, S-complex asteroids have a trend of the medium to steep slope. 
For S-type asteroids, it is known that the space weathering causes a significant change in vis-NIR spectra, 
in which the albedo decreases (the spectrum becomes darker) and the spectral slope increases (redder)
\citep{Yamada1999}. 
For the space weathering of C-complex asteroids, many laboratory experiments 
have been conducted (for recent studies, e.g., \citealt{Matsuoka2017, Lantz2017}). 
However, it is still in debate whether 
the space weathering makes the spectral slope of 
asteroids other than S-type bluer or redder, and darker 
or brighter. 

\subsection{From dry to wet: characteristics of hydrated asteroids}

In the field of meteorite study, it is widely agreed that CM, CI, and CR carbonaceous chondrites, 
which originate in C-complex asteroids,  experienced aqueous alteration in their parent bodies 
(\citealt{Brearley2006} and references cited therein).
This aqueous alteration occurred from the reaction of anhydrous rock and liquid water with the isotope decay heat 
(\citealt{McAdam2015} and references cited therein). Correspondingly, 
from the standpoint of astronomical observations, hydration in asteroids has been 
discussed with a 0.7-$\micron$ absorption feature; C-complex asteroids with an absorption feature in 
the broad 0.7-$\micron$ band are categorized as Ch- or Cgh-type \citep{Bus2002}, where suffix ``h'' represents 
a hydrated subclass (e.g., \citealt{Burbine1993}). This 0.7-$\micron$ feature is indicative of
an Fe$^{2+}$~$\to$~Fe$^{3+}$ charge transfer transition in oxidized iron in phyllosilicates formed 
through aqueous alteration processes (e.g., \citealt{Vilas1989}). On the other hand, absorption in 
the 2.7-$\micron$ band is attributed to OH-stretch in hydrated minerals and is a more direct indicator 
of the presence of water. 

\begin{figure}
\begin{center}
\includegraphics[width=140mm]{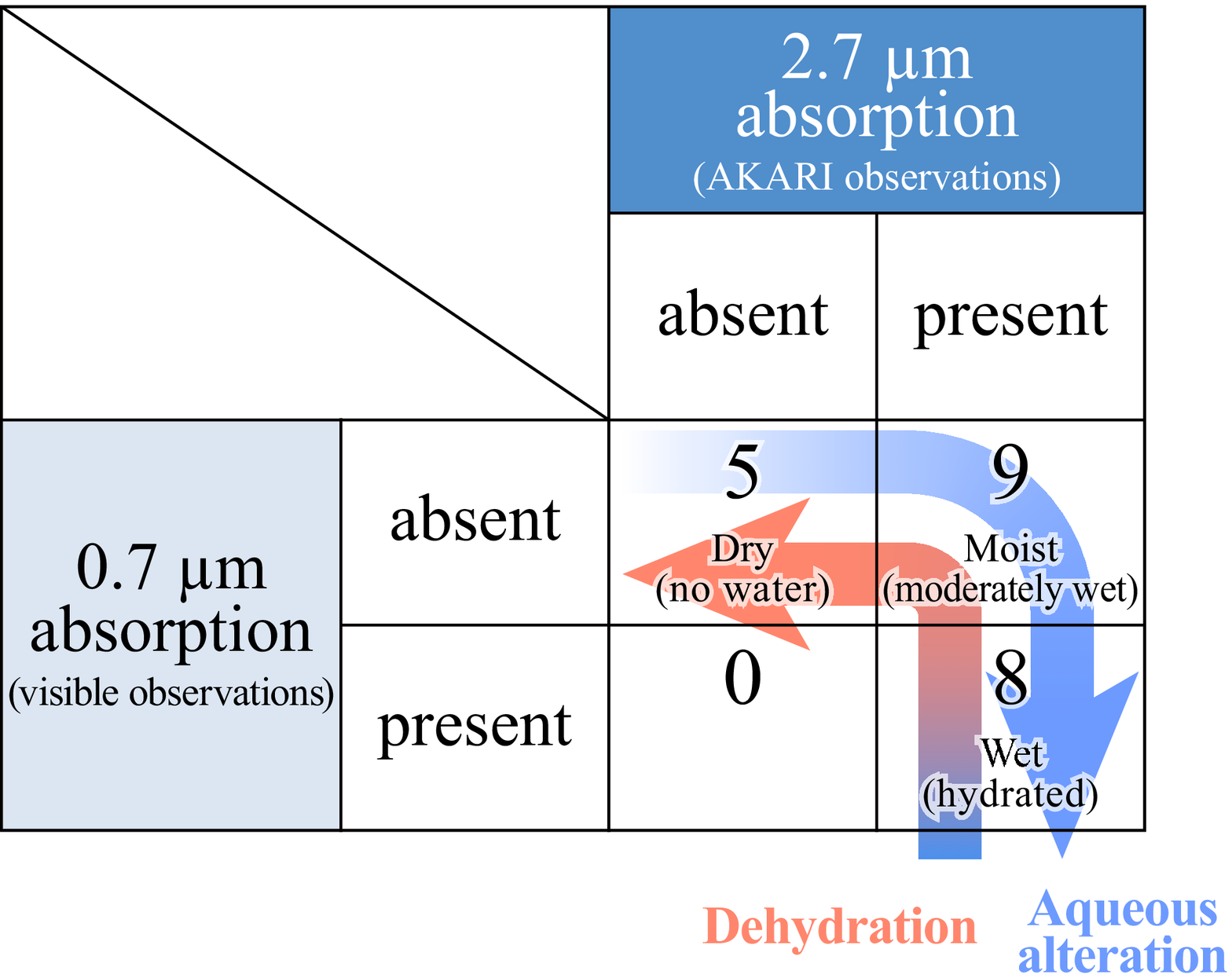}
\end{center}
\caption{
Matrix of a relation between the 0.7- and 2.7-$\micron$ band features in all C-complex asteroids 
observed in this study.}\label{fig:0.7-2.7um} 
\end{figure}

\begin{figure}
\begin{center}
\includegraphics[width=140mm]{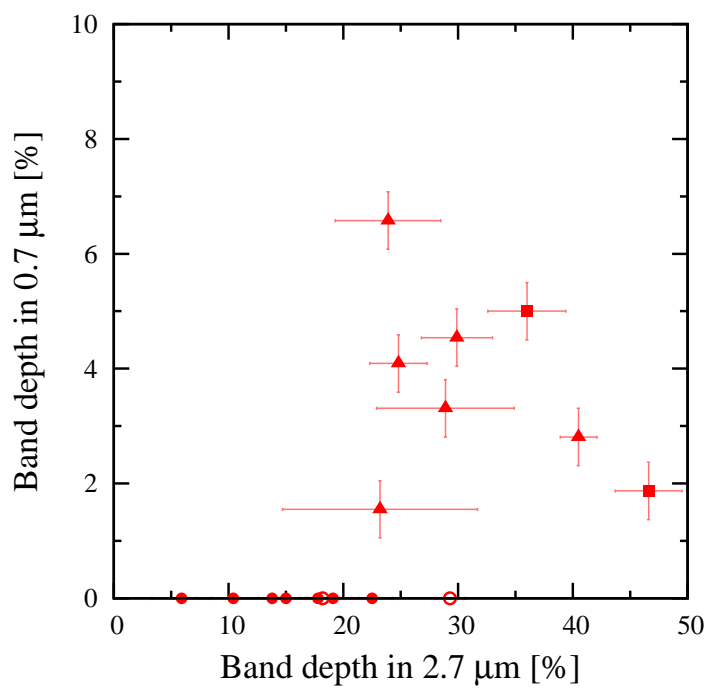}
\end{center}
\caption{
Correlation between the 2.7- and 0.7-$\micron$ band depths in all C-complex asteroids 
observed in this study.}\label{fig:0.7-2.7um strength} 
\end{figure}

Figure~\ref{fig:0.7-2.7um} shows the number of C-complex asteroids observed in the 0.7- and 
2.7-$\micron$ band in this study (c.f., \citealt{Rivkin2002,Rivkin2015AIV}). 
Figure~\ref{fig:0.7-2.7um strength} shows the distribution of band depths in the 0.7- and 2.7-$\micron$ 
bands. 
The 0.7-$\micron$ band depth is measured from the visible spectra compiled by \cite{Hasegawa2017}. 
\cite{2015AJ....150..198R} discussed the correlation presented in the distribution of 
the 0.7- and 2.7-$\micron$ band depths based on the combined set of asteroid and meteorite measurements. 
No clear trends were found in figure~\ref{fig:0.7-2.7um strength}, 
partly due to a lack of adequate samples for the statistical study. 
\cite{Fornasier2014} pointed out the relationship between presence/absence of the 0.7- and 2.7-$\micron$ 
band features as; if the 0.7-$\micron$ band is present on the spectrum, it is always accompanied by 
the 2.7-$\micron$ feature. On the other hand, even if the 0.7-$\micron$ band is not found, the 2.7-$\micron$ 
band may still be present on the spectrum. In our measurements, the 0.7-$\micron$ band depth is generally 
weaker by a few percent than that of the 2.7-$\micron$ band. Our results confirm those of \cite{Fornasier2014}; 
in addition, not a small number of objects are found without both the 0.7- and 2.7-$\micron$ band 
features. \cite{Rivkin2012b} reported that 60-70\% of C-complex asteroids are expected to have hydrated 
minerals. Our results also indicate that 77\% of the observed C-complex asteroids have the 2.7-$\micron$ 
feature, although it may suffer from observational bias. 

To discuss observational signatures with the aqueous alteration, we classify 
asteroids into three stages in this study as: 
\begin{itemize}
\item ``dry'' objects have no feature both in the 0.7- and {2.7-$\micron$} bands, 
\item  ``moist'' objects have feature only in the 2.7-$\micron$ band, 
\item ``wet'' objects have features both in the 0.7- and 2.7-$\micron$ bands. 
\end{itemize}
It should be stressed that 
the terms ``moist'' and ``wet'' are figurative and not literal meaning; 
it does not mean the presence of water in the form of vapor or liquid, but 
describe the relative amount of H$_2$O or OH in minerals. 
Note that water vapor was detected on 1~Ceres (classified into ``moist'') 
with the Herschel Space Observatory \citep{Kuppers2014}. 
Figure~\ref{fig:0.7-2.7um} can be interpreted in terms of the sequence of 
aqueous alteration and dehydration process on asteroids with the three stages described above 
as: 
\begin{enumerate}
\item An asteroid is ``dry'' if the signature of aqueous alteration does not appear on 
its surface --- it may be formed in an environment without the presence of water, i.e., dry condition. 
\item As aqueous alteration progresses to a certain degree, it appears as the 2.7-$\micron$ 
band feature. It becomes a ``moist'' (or moderately wet) object. 
\item When aqueous alteration progresses sufficiently, the spectrum shows the 0.7-$\micron$ 
feature and the object becomes ``wet''. 
The progress of aqueous alteration does not necessarily progress completely to this stage, and 
the reaction may stop halfway without the appearance of the 0.7-$\micron$ feature. 
\item Dehydration occurs by solar-radiation heat or impact-induced heat. As dehydration 
progresses, the 0.7-$\micron$ feature disappears 
and the asteroid changes from the ``wet'' to the ``moist'' stage.
\item As dehydration progresses further, the 2.7-$\micron$ feature also disappears and 
the object returns to the ``dry'' stage.
\end{enumerate}
Let it be added, in this context, that a certain amount of evidence from meteorite study 
suggest the aqueous alteration of carbonaceous chondrites (e.g., \citealt{Brearley2006}). 
These carbonaceous chondrites are considered to originate from C-complex asteroids, or more specifically, 
formed inside C-complex asteroids (e.g., \citealt{McSween1999}). 
In other words, hydrated minerals in C-complex asteroids, which show the 2.7-$\micron$ band feature, 
are considered to be 
of an internal (endogenic) origin. 
This suggests that C-complex asteroids were formed in the environment where 
anhydrous rock and water ice existed together at low temperature 
in the protoplanetary disk at the time of planetesimal formation. 
On the other hand, the degree of aqueous alteration in ordinary chondrites is found to be relatively small 
(e.g., \citealt{Brearley2006}). 
It is consistent with that 
only a few S-complex asteroids show signatures of aqueous alteration in this study. 

The situation of X- and D-complex asteroids is more complicated. 
The major reason why interpretation of these asteroids is difficult is that 
there are few spectral counterparts of meteorites found to X- or D-complex asteroids 
(c.f., \citealt{2003GeoRL..30.1909K, HiroiHasegawa2003}). 
Nevertheless, it can be considered that low-albedo X-complex (i.e., P-type) and D-complex asteroids 
may possess properties similar to C-complex 
in the sense of primitive objects (e.g., \citealt{Vernazza2015}). 
Thus it is natural that some of these asteroids have signatures of hydration 
as seen in some stages of C-complex asteroids as the 2.7-$\micron$ band feature in this study 
(e.g., 56~Melete, 476~Hedwig, and 308~Polyxo). 

There is an exceptional case found from our results that 349~Dembowska 
(R-type; olivine-rich, basaltic asteroid, \citealt{Leith2017}) 
has a 2.7-$\micron$ band feature with 4\%. It is considered to originate from external 
(exogenic) materials, which was brought to its surface by hydrated impactors 
or created by solar wind interactions with silicates.  
Recent studies also 
reported that exogenic materials are detected in such traditionally dry asteroids, 
for example, 4~Vesta (V-type; \citealt{McCord2012, 2012ApJ...758L..36D}), 
16~Psyche (M-type: \citealt{2017AJ....153...31T, Avdellidou2018}), 
and 433~Eros and 1036~Ganymed (S-type; \citealt{Rivkin2018}). 
This ``contamination'' of exogenic materials may be not rare in the main-belt region. 


\section{Summary}

We conducted a near-infrared spectroscopic survey with the AKARI satellite
to obtain reflectance spectra of 66 asteroids in the 2.5--5~$\micron$ range. 
These observations successfully fill the gap in the 2.5--2.85~$\micron$ data, which cannot 
be observed with ground-based telescopes. 
Based on the spectra obtained with AKARI, we found that most C-complex asteroids 
have clear absorption features related to hydrated minerals at a peak wavelength of approximately 
2.75~$\micron$, while no S-complex asteroids have clear absorption in this wavelength. 

This data set, comprising direct observations of absorption features in the 2.7-$\micron$ band, is quite unique. 
This study will provide important information on whether asteroid features determined by spacecraft exploration are universal or exceptional. 
Combining this data set with spectra in shorter wavelengths ($< 2.5~\micron$), and comparing it to meteorite spectra 
measured in the laboratory are both interesting research subjects. These will be discussed at length in future papers. 
Our spectral data are summarized in the Asteroid Catalog using AKARI Spectroscopic Observations (AcuA-spec) 
and open to the public on the JAXA archive\footnote{http://www.ir.isas.jaxa.jp/AKARI/Archive/}. 

\begin{ack}
This work is based on observations with AKARI, a JAXA project with the participation of ESA. 
This study was supported by JSPS KAKENHI: Grant Numbers JP15K05277, JP17K05381, and JP17K05636,
and partly by the Center of Planetary Science, Kobe University, and the Hypervelocity Impact Facility 
(former facility name: the Space Plasma Laboratory), ISAS/JAXA. 
This study utilized the JPL HORIZONS ephemeris generator system, operated at JPL, Pasadena, USA, 
and Lowell Asteroids Services, operated at Lowell Observatory, Flagstaff, USA. 
A pilot survey of this work was conducted by Natsuko Okamura and Seiji Sugita (The University of Tokyo). 
The authors greatly appreciate 
Takahiro Hiroi (Brown University), 
Ellen S. Howell (University of Arizona), 
Zoe A. Landsman (University of Central Florida), 
Javier Licandro (Instituto de Astrofisica de Canarias), 
Andrew S. Rivkin (The Johns Hopkins University Applied Physics Laboratory), 
Driss Takir (SETI Institute), Pierre Vernazza (Laboratoire d'Astrophysique de Marseille), 
and 
Makoto Yoshikawa (ISAS/JAXA), 
for kindly providing their data. 
FU would like to acknowledge 
Joshua P. Emery (University of Tennessee), 
Thomas G. M\"uller, Victor Al\'i-Lagoa (Max-Planck-Institut f\"{u}r Extraterrestrische Physik), 
Daisuke Kuroda (Kyoto University), 
Tomoki Nakamura (Tohoku University), 
Takaaki Noguchi (Kyushu University), 
Sei-ichiro Watanabe (Nagoya University), 
Takao Nakagawa, Takehiko Wada, Mitsuyoshi Yamagishi (ISAS/JAXA), and 
Takayuki Ushikubo (Japan Agency for Marine-Earth Science and Technology)  
for their helpful comments and discussions. 
Finally, we are very grateful to the anonymous reviewer for careful reading and providing constructive suggestions. 

\end{ack}

\appendix

\section{References of previous studies in the 3-$\micron$ band}
\label{reference of previous studies}
Reference list of previous spectroscopic observation research in the 3-$\micron$ band 
are given in table \ref{tab:previous_works}. 

\clearpage
\onecolumn

\clearpage

\clearpage

\clearpage

\clearpage

\clearpage

\clearpage

\clearpage

\clearpage

\clearpage

\clearpage

\clearpage

\clearpage

\clearpage

\clearpage

\clearpage

\clearpage

\clearpage

\clearpage

\clearpage




\begin{table}
\caption{Candidate of groupings for objects in the 3-$\micron$ band shape.}\label{tab:3um group}
%

\end{document}